\documentclass[12pt,draftclsnofoot,onecolumn]{IEEEtran}
\usepackage{tabularx}
\usepackage[T1]{fontenc}% optional T1 font encoding
\usepackage{algorithmic}
\usepackage{graphicx}
\usepackage{amsmath,amssymb,amsfonts}
\usepackage{array}
\usepackage{textcomp}
\usepackage[font=small]{caption}
\usepackage{paralist}
\usepackage{lettrine}
\usepackage{cite}
\usepackage{xcolor}
\usepackage{cancel}
\usepackage{float}
\usepackage{booktabs}
\usepackage{subfig}
\usepackage{stfloats}
\usepackage{url}
\usepackage{lipsum}
\usepackage{enumitem}
\usepackage{mathtools}
\usepackage{cuted}
\usepackage{makecell}
\usepackage{acronym}
\usepackage{multicol}
\usepackage{multirow}
\usepackage[ruled,vlined]{algorithm2e}
\usepackage{authblk}
\usepackage{fixltx2e}

\usepackage[margin=0.93in]{geometry}

\newcommand{\pdv}[2]{\frac{\partial #2}{\partial #1}}

\newcommand{\RN}[1]{%
  \textup{\uppercase\expandafter{\romannumeral#1}}%
}

\def\BibTeX{{\rm B\kern-.05em{\sc i\kern-.025em b}\kern-.08em
    T\kern-.1667em\lower.7ex\hbox{E}\kern-.125emX}}

\title{Conformal Metasurfaces: \\ a Novel Solution for Vehicular Communications}

\author{Marouan Mizmizi, Reza Aghazadeh Ayoubi, Dario Tagliaferri, Kai Dong \\ Gian Guido Gentili, and Umberto Spagnolini}

\begin{document}
\maketitle

\begin{abstract}
In future 6G millimeter wave (mmWave)/sub-THz vehicle-to-everything (V2X) communication systems, vehicles are expected to be equipped with massive antenna arrays to realize beam-based links capable of compensating for the severe path loss. However, vehicle-to-vehicle (V2V) direct links are prone to be blocked by surrounding vehicles. Emerging metasurface technologies enable the control of the electromagnetic wave reflection towards the desired direction, enriching the channel scattering to boost communication performance. Reconfigurable intelligent surfaces (RIS), and mostly the pre-configured counterpart intelligent reflecting surfaces (IRS), are a promising low-cost relaying system for 6G. This paper proposes using conformal metasurfaces (either C-RIS or C-IRS) deployed on vehicles' body to mitigate the blockage impact in a highway multi-lane scenario. In particular, conformal metasurfaces create artificial reflections to mitigate blockage by compensating for the non-flat shape of vehicle's body, such as the lateral doors, with proper phase patterns. We analytically derive the phase pattern to apply to a cylindrical C-RIS/C-IRS approximating the shape of car body, as a function of both incidence and reflection angles, considering cylindrical RIS/IRS as a generalization of conventional planar ones. We propose a novel design for optimally pre-configured C-IRS to mimic the behavior of an EM flat surface on car doors, proving the benefits of C-RIS and C-IRS in a multi-lane V2V highway scenario. The results show a consistent reduction of blockage probability when exploiting C-RIS/C-IRS, $20\%$ for pre-configured C-IRS and $70\%$ for C-RIS and, as well as a remarkable improvement in terms of average signal-to-noise ratio, respectively $10-20$ dB for C-IRS and $30-40$ dB for C-RIS. 

\end{abstract}

\section{Introduction}

Road mobility is experiencing an unprecedented technological transformation towards safer and efficient vehicular networks. Vehicle-to-everything (V2X) communication and autonomous guidance technologies are the main engines driving innovation and the development of novel services for road users. V2X technology allows the data sharing among vehicles, i.e., vehicle-to-vehicle (V2V), with infrastructure in vehicle-to-infrastructure (V2I) or vehicle-to-network (V2N), and with pedestrians in vehicle-to-pedestrian. To support V2X technology, the third generation partnership project (3GPP) has introduced cellular-V2X (C-V2X) in its Release 14 \cite{3GPPTR22185} and further enhancements in Releases 15 and 16 for the 5G V2X \cite{3GPPTR22886-R16}, operating at sub-6GHz frequencies. However, due to the limited bandwidth, the current 5G V2X and C-V2X cannot support the requirements of advanced V2X services \cite{3GPPTR22186}. 

In recent years, 3GPP has proposed the use of millimeter-wave (mmWave) frequencies ($30-100$ GHz) in the next release 17 \cite{3GPPTR38901, harounabadi2021v2x} for V2X communications, while sub-THz frequencies ($>100$ GHz) are being considered for the upcoming 6G systems \cite{rappaport2019wireless, tripathi2021millimeter}. The propagation at these frequencies presents several technological challenges. Path loss and penetration loss are orders of magnitude higher than current systems \cite{jameel2018propagation}, resulting in limited covered range and frequent communication drop-out due to blockage \cite{tunc2019millimeter}, which is harsher in dynamic scenarios such as in the vehicular one. 

One of the main source of blockage in V2V scenarios are cars interposing between transmitting vehicle (TxV) and receiving vehicle (RxV). At $30$ GHz, this can cause a power loss ranging from $10$ to $20$ dB \cite{park2017millimeter}, depending on the relative blocker's position, and increases by $5.5$ to $17$ dB when multiple cars are simultaneously blocking the line-of-sight (LoS). In our previous work \cite{dong2021vehicular}, we derived the analytical blockage probability due to multiple vehicles, showing a severe impact on the V2V link with increasing TxV-RxV distance and/or traffic intensity.
Hence, blockage mitigation solutions are of utmost importance to guarantee the required robustness and reliability of mmWave and sub-THz links. The work in \cite{linsalata2021map} proposes a method for blockage prediction leveraging on active relays of opportunity. However, active relays need to be scheduled and require spare resources dedicated to relaying. The authors in \cite{Panwar8643739} derived an analytical model of the blockage and propose to use the macro-diversity of base stations to reduce the probability of link interruption, suggesting the increase of base stations' number at the price of much higher deployment costs.

A different class of strategies leverages on the emerging paradigm of smart radio environment (SRE)~\cite{di2020smart, sun2019electromagnetic}, which aims at enriching the scattering environment between transmitter and receiver. The radio propagation channel is controlled using intelligent surfaces (or metamaterials) to reflect the impinging radio waves in specific directions. These surfaces are made by nearly passive arrays of sub-wavelength sized elements whose complex reflection coefficient can be tuned (either amplitude and phase, or phase only) to manipulate the impinging wavefront and control the reflection angle~\cite{zhang2021performance}. Herein, we refer to reconfigurable intelligent surfaces (RIS) when the amplitude and phase of each element can be set in real-time \cite{abeywickrama2020intelligent} and to intelligent reflecting surfaces (IRS) when it is pre-configured and not tunable after manufacturing \cite{oliveri2021holographic}. Although the RIS and IRS terminologies are often used interchangeably in literature, we herein stress this distinction to distinguish between fully passive surfaces (IRS) and nearly passive ones (RIS). Moreover, RIS are more flexible in dynamic environments but require dedicated control signaling, while IRS are cheaper and need no power supply and control signaling.

In particular, RIS have been largely investigated in literature for improving the average capacity per unit area \cite{direnzo2021surveyRIS} and to address the blockage issue in vehicular networks. Referring to blockage management, in \cite{Zeng9539048}, authors propose a RIS-assisted handover scheme using deep reinforcement learning to mitigate the blockage in a cellular scenario. In the considered setting, the base station computes the RIS configuration for each cluster of users based on the observed channels. The proposed approach does not explicitly solve a non-convex optimization problem for RIS configuration as in \cite{chen2020resource, chen2021qos}, but still requires additional signaling overhead, which might be not suited to a highly dynamic V2V scenario. Although some works consider a finite phase set at each RIS element to decrease the control overhead \cite{shao2019dielectric, yin2020single}, the real-time RIS reconfiguration is still an open issue. The problem of RIS deployment optimization to support V2I/V2N communication is investigated in \cite{Kurt9359529}, showing a remarkable gain in received power when multiple RIS are strategically deployed. However, the overall deployment cost and signaling overhead for configuration (not addressed in the paper) drastically increase with the number of RIS with questionable economic impact. 
Most works assume perfect channel state information (CSI) is available at the RIS, or they consider slowly varying CSI scenarios \cite{Zhou9110587}. However, for V2V links, the CSI outdates rapidly \cite{ashraf2020supporting, garcia2021tutorial}, making the state-of-the-art solutions inefficient or even impractical.
By contrast, IRS can offer a fully passive alternative to RIS when either the deployment cost is excessive or the incidence/reflection angles are not a-priori known and cannot be properly estimated. For instance, IRS can be deployed on buildings to enrich the environment scattering, improving the coverage and the average communication performance over different users in different positions in space~\cite{oliveri2021holographic,benoni2021planning}.
However, the density of building-deployed metasurfaces required for coverage might be critical in highly urbanized areas, especially to support V2V communications at the road level.

\subsection*{Contributions}

This paper aims at augmenting the SRE with conformal metasurfaces mounted on (or hidden on) cars' body. To the best of the authors' knowledge, no work has considered directly equipping vehicles with metasurfaces. In this context, this paper proposes to mount conformal metasurfaces (C-RIS/C-IRS) on both sides of each car (i.e., on both curved surfaces of the lateral doors) to act as nearly passive or fully passive relays of opportunity for blocked V2V links. The core idea is that future connected and automated vehicles (CAVs) will be equipped with antenna arrays for V2X communications and can be augmented with C-RIS/C-IRS on their sides.
In all the aforementioned works, RIS/IRS were planar arrays of reconfigurable/pre-configured elements mounted on flat surfaces only. Car doors are, in general, conformal surfaces that cannot be considered as EM flat at mmWave/sub-THz frequencies. Conformal metasurfaces have been investigated in \cite{li2020arbitrarily, la2019curvilinear, qian2020deep} for acoustic wave and light manipulation, but never at radio frequency and for vehicular applications.
Hence, in this paper, we first analytically derive the phase pattern for cylindrical C-RIS, which can be regarded as an approximation of the complex non-planar shape of car doors. We stem from the generalized reflection law for arbitrarily shaped surfaces, where both the incidence and reflection angles are perfectly known, i.e., perfect CSI. Then, to avoid the explicit real-time reconfiguration of C-RIS and the associated control signaling overhead, we propose a novel pre-configured (or static) phase pattern design for fully passive C-IRS. The aim is to compensate for the curvature of the car doors (or any conformal surface), mimicking the behavior of a perfectly flat surface. Herein, C-IRS is designed to provide a significant specular reflection without requiring that other vehicles waste their radio resources (and energy) for relaying, in a win-win model where all vehicles will benefit from this passive relaying.
Blockage probability reduction and average signal-to-noise ratio (SNR) gains are demonstrated in a multi-lane highway scenario. In particular, the combined usage of C-RIS and direct V2V links yield a reduction of $70\%$ in blockage probability and approximately $30$ dB SNR gain compared to the direct V2V link only. Using fully passive pre-configured C-IRS leads to lower benefits, namely $-20\%$ in blockage probability and $15-20$ dB of SNR gain. However, C-IRS does not require the CSI knowledge as for C-RIS; hence no control signaling is required. In addition, we also provide some insights for the practical realization of fully passive C-IRS, easing the massive implementation in 6G vehicular networks. 

\subsection*{Organization}
The remainder of the paper is organized as follows: Section \ref{sect:system_model} describes the system and channel model and Section \ref{sect:reflection_conformal} outlines the proposed conformal metasurface design. Section \ref{sect:numerical_results} validates the remarkable benefits from the adoption of C-IRS for a V2V scenario. Finally, conclusions and open challenges are in Section \ref{sect:conclusion}.

\subsection*{Notation}
Bold upper- and lower-case letters describe matrices and column vectors. The $(i,j)$-th entry of matrix $\mathbf{A}$ is denoted by $[\mathbf{A}]_{(i,j)}$. Matrix transposition, conjugation, conjugate transposition and Frobenius norm are indicated respectively as $\mathbf{A}^{\mathrm{T}}$, $\mathbf{A}^{*}$, $\mathbf{A}^{\mathrm{H}}$ and $\|\mathbf{A}\|_F$. $\mathrm{tr}\left(\mathbf{A}\right)$ extracts the trace of $\mathbf{A}$. $\mathrm{diag}(\mathbf{A})$ denotes the extraction of the diagonal of $\mathbf{A}$, while $\mathrm{diag}(\mathbf{a})$ is the diagonal matrix given by vector $\mathbf{a}$. $\mathbf{I}_n$ is the identity matrix of size $n$. With  $\mathbf{a}\sim\mathcal{CN}(\boldsymbol{\mu},\mathbf{C})$ we denote a multi-variate circularly complex Gaussian random variable $\mathbf{a}$ with mean $\boldsymbol{\mu}$ and covariance $\mathbf{C}$. $\mathbb{E}[\cdot]$ is the expectation operator, while $\mathbb{R}$ and $\mathbb{C}$ stand for the set of real and complex numbers, respectively. $\delta_{n}$ is the Kronecker delta.

\section{System and Channel Model}\label{sect:system_model}

\begin{figure}[b!]
    \centering
    \includegraphics[width=0.45\textwidth]{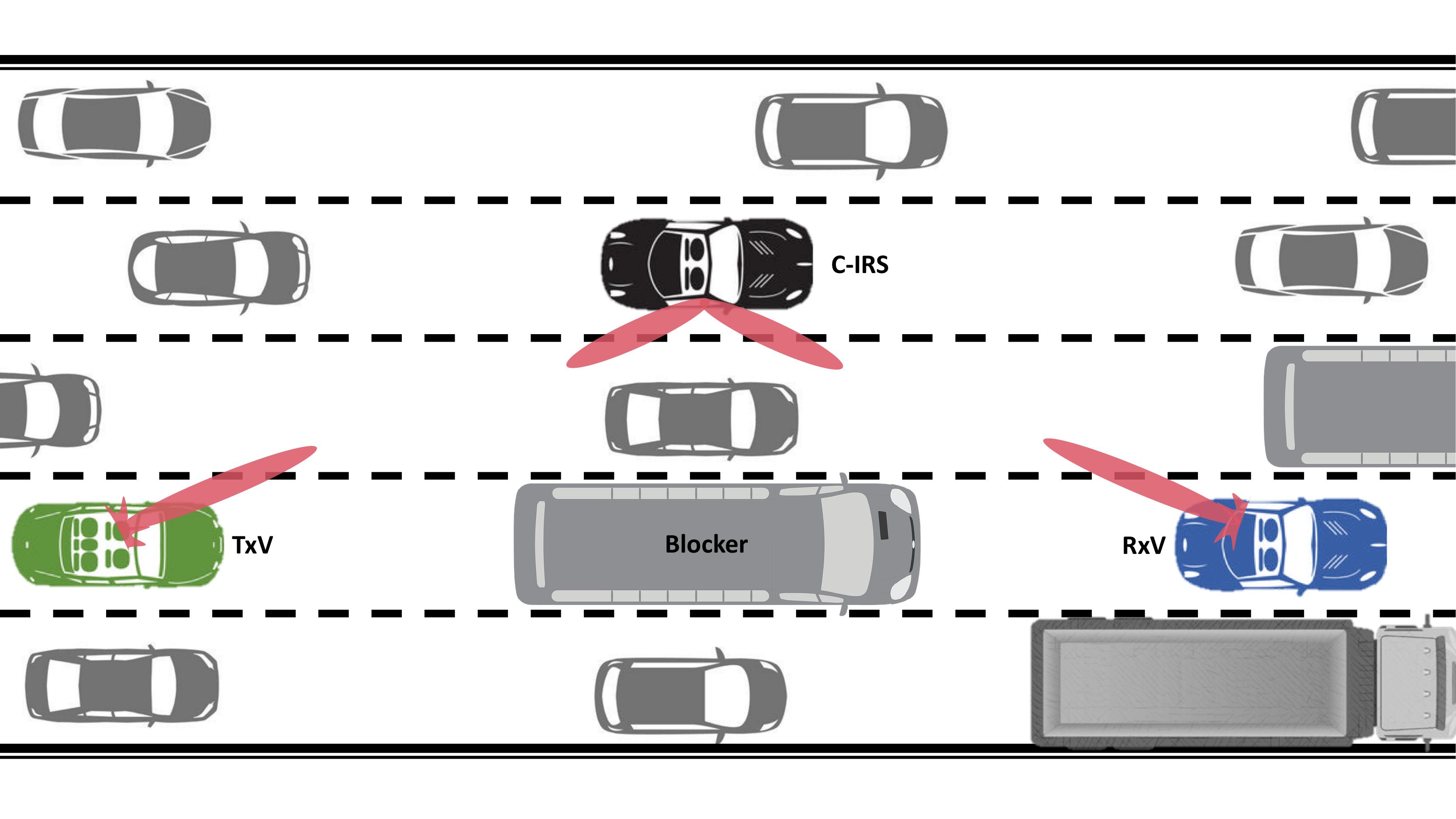}
    \caption{Multi-lane highway scenario: the direct path between TxV and RxV might be blocked by other vehicles, thus other cars in other lanes can act as passive relays}
    \label{fig:scenario}
\end{figure}
\begin{figure*}
    \centering
    \includegraphics[width=0.95\textwidth]{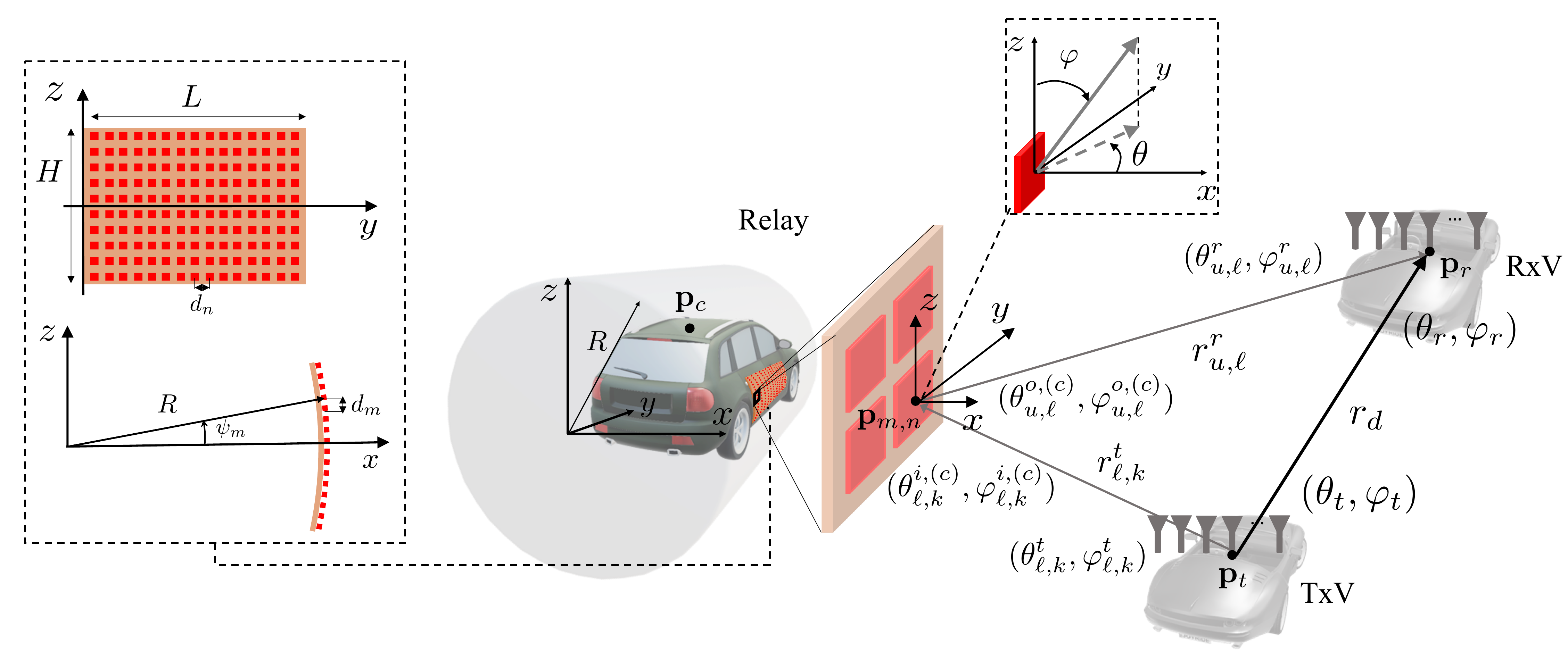}
    \caption{Sketch of the geometry and the reference system of the considered V2V scenario. Each vehicle is equipped with a C-RIS/C-IRS on each side to serve as an opportunistic relay. The conformal door shape is approximated with a cylinder of radius $R$, in which each element of the C-RIS/C-IRS along the conformal coordinate is identified by angle $\psi_m$.}
    \label{fig:systemModel}
\end{figure*}

We consider the multi-lane highway vehicular scenario depicted in Fig. \ref{fig:scenario}. 
Each vehicle is equipped with a uniform linear array (ULA) with $K$ antenna elements, and two conformal metasurfaces (C-RIS/C-IRS) on both vehicle's sides. 
At a given time instant, the position of TxV and RxV antenna arrays is $\mathbf{p}_{t} = [x_{t}, y_{t}, z_{t}]^\mathrm{T}$ and $\mathbf{p}_{r} = [x_{r}, y_{r}, z_{r}]^\mathrm{T}$, respectively, defined in a global coordinate system. Similarly, the relaying vehicle is located in $\mathbf{p}_{c} = [x_{c}, y_{c}, z_{c}]^\mathrm{T}$, where $\mathbf{p}_{c}$ identifies the position of a reference element of the C-RIS/C-IRS. The global coordinate system is such that the cars move along direction $y$, the cross-motion axis is $x$ and $z$ denotes the vertical direction.

Herein, the shape of each C-RIS/C-IRS along the door is assumed cylindrical with a curvature radius $R$. In general, a car door has an arbitrarily complex shape; however, the cylindrical assumption allows for a closed-form analytical solution and to generalize the solution of planar RIS/IRS ($R \rightarrow \infty$).
In this setting, $M$ elements of the cylindrical metasurface are deployed along the curved coordinate (height), and $N$ elements along the cylindrical direction (length), for a total of $M\times N$ elements. The equipment is sketched in~Fig. \ref{fig:systemModel}, where the length of the car's door is $L$ and the height is $H$. The position of the $(m,n)$-th C-RIS/C-IRS element, for $m = -M/2, \dots, M/2-1$ and $n = 0,\dots, N-1$, can be expressed in global coordinates as $\mathbf{p}_{m,n} = \mathbf{p}_c + [x_{m,n},\,y_{m,n},\,z_{m,n}]^\mathrm{T}$, where $x_{m,n}$, $y_{m,n}$ and $z_{m,n}$ denote the relative 3D displacement of the $(m,n)$-th element with respect to the reference one. Here, $x_{m,n} = R (\cos \psi_m-1)$, $y_{m,n} = d_n (n-1)$ and $z_{m,n} = R \sin\psi_m$, where $\psi_m = m 2\arcsin\left(d_m/2/R\right)$ is the angular position in cylindrical coordinates of the $m$-th row, while $d_m$ and $d_n$ are the elements' spacing along the vertical and horizontal directions, respectively.
The area of the C-RIS/C-IRS is therefore:
\begin{equation}
    A = L \times 2R \,\psi_{M}
\end{equation}
where $\psi_{M}=M\arcsin\left(d_m/2/R\right)$ is the angular sector spanned by the C-RIS/C-IRS. For a reference car door height and length of $H=1$ m and $L=1$ m, selecting $R\in[1,8]$ m provides a curvature that is in line with common curved car doors \cite{DoorRadius}. Notice that, $\lambda=1.07$ cm, i.e., $28$ GHz carrier frequency, and $d_m=\lambda/4$, the C-RIS/C-IRS shall be composed by $M\approx 380$ elements along the curved direction.   

\subsection{Signal Model}\label{subsect:signal_model}

Let the complex symbol transmitted be $s \in \mathbb{C} \sim \mathcal{CN}\left(0, \sigma_s^2\right)$, where $\sigma_s^2$ is the transmitted power. Symbol $s$ is beamformed $\mathbf{f} \in \mathbb{C}^{K \times 1}$, and the transmitted signal is 
\begin{equation}
    \mathbf{x} = \mathbf{f}\,s.
\end{equation}
The propagation is over a block-faded spatially-sparse channel $\mathbf{H} \in \mathbb{C}^{K \times K}$, whose model is detailed in the next subsection.
After the time-frequency synchronization, the received symbol $y\in \mathbb{C}$ is
\begin{equation}\label{eq:receivedSignal}
    y = \mathbf{w}^\mathrm{H} \mathbf{H} \mathbf{f} \,s + \mathbf{w}^\mathrm{H} \mathbf{n},
\end{equation}
where $\mathbf{w} \in \mathbb{C}^{K \times 1}$ is the beamformer (combiner) at RxV, and the additive noise $\mathbf{n} \sim \mathcal{CN}\left(\mathbf{0}, \mathbf{Q}_n\right)$ have a covariance $\mathbf{Q}_n = \mathbb{E}\left[\mathbf{nn}^\mathrm{H}\right]$ that for simplicity is $\mathbf{Q}_n=\sigma^2_n\mathbf{I}_K$.
Derivation of beamformers $\mathbf{f}$ and $\mathbf{w}$ is detailed in Section \ref{sect:numerical_results}.

\subsection{Channel model}\label{subsect:channel_model}

The channel matrix in \eqref{eq:receivedSignal} is the sum of two contributions \cite{Sanguinetti9300189}:
\begin{equation}\label{eq:channelDefinition}
    \mathbf{H} = \mathbf{H}_d + \mathbf{H}_{cr} \boldsymbol{\Phi} \mathbf{H}_{tc}
\end{equation}
where the first term $\mathbf{H}_d$ is the direct TxV-RxV link and the second term is via the C-IRS. Specifically, $\mathbf{H}_{tc} \in \mathbb{C}^{MN \times K}$ is the TxV-C-IRS channel for the incident signal, $\mathbf{H}_{cr} \in \mathbb{C}^{K \times MN}$ is the C-IRS-RxV channel for the reflected signal, and $\boldsymbol{\Phi}\in \mathbb{C}^{MN \times MN}$ is the complex reflection matrix (amplitude and phase) of the C-IRS (C-RIS). 

As customary in mmWave/sub-THz communications, the channel \eqref{eq:channelDefinition} exhibits a sparse scattering characteristic \cite{rappaport2019wireless}. The model for the direct channel $\mathbf{H}_d$ can be written for the far-field assumption, that tipically applies for TxV-RxV distances in the order of tens of meters, as:
\begin{equation}\label{eq:channelModel}
    \mathbf{H}_d = \alpha_d\, \varrho_r(\boldsymbol{\vartheta}_d^r) \varrho_t(\boldsymbol{\vartheta}_d^t) \mathbf{a}_r(\boldsymbol{\vartheta}_d^r)\mathbf{a}_t(\boldsymbol{\vartheta}_d^t)^\mathrm{H} 
\end{equation}
where \textit{(i)} $\alpha_d$ denotes the complex gain of the direct path, \textit{(ii)} $\mathbf{a}_t(\boldsymbol{\vartheta}_d^t)\in\mathbb{C}^{K\times 1}$, $\mathbf{a}_r(\boldsymbol{\vartheta}_d^r)\in\mathbb{C}^{K\times 1}$ are the TxV and RxV array response vectors, function of the angles of arrival (AoAs) and angles of departure (AoDs) of the direct path, respectively $\boldsymbol{\vartheta}_d^r = (\theta_d^r, \varphi_d^r)$ and $\boldsymbol{\vartheta}_d^t = (\theta_d^t, \varphi_d^t)$ (for azimuth and elevation), \textit{(iv)} $\varrho_t(\boldsymbol{\vartheta}_d^t)$ and $\varrho_t(\boldsymbol{\vartheta}_d^r)$ are the TxV and RxV single-antenna gains, respectively. Herein, we consider two half-wavelength spaced ULAs for both TxV and RxV, the array response at center bandwidth being:
\begin{equation}\label{eq:beamformer}
    \mathbf{a}(\theta) = \frac{1}{\sqrt{K}}\left[1,...,e^{-j\pi(K-1)\cos\theta}\right]^\mathrm{T} 
\end{equation}
While the far-field assumption generally holds for the direct TxV-RxV paths, the link through the C-IRS can be either in far-field (with planar wavefront) or near-field (with curved wavefront) depending on the size of the C-IRS compared to the TxV-IRS and IRS-RxV distances. Therefore, the $(\ell,k)$-th entry of channel matrices $\mathbf{H}_{tc}$ and $\mathbf{H}_{cr}$, for $\ell = (m-1)M+n$, $k=1,...,K$ is ,
\begin{equation}\label{eq:generalModel}
\begin{split}
    &\left[\mathbf{H}_{tc}\right]_{(\ell,k)} = \alpha_{\ell,k}\; \varrho_t(\boldsymbol{\vartheta}_{\ell,k}^t) \, \varrho_c(\boldsymbol{\vartheta}_{\ell,k}^{i,(c)}) \, e^{-j\frac{2\pi}{\lambda}r_{\ell,k}^t}\\
    &\left[\mathbf{H}_{cr}\right]_{(u,\ell)} = \alpha_{u,\ell}\;
    \varrho_c(\boldsymbol{\vartheta}_{u,\ell}^{o,(c)})
    \varrho_r(\boldsymbol{\vartheta}_{u,\ell}^r) \, e^{-j\frac{2\pi}{\lambda}r_{u,\ell}^r}
\end{split}
\end{equation}
where \textit{(i)} $\alpha_{\ell,k}$ and $\alpha_{u,\ell}$ are the complex gains of the path between the $k$-th TxV antenna and the $\ell$-th C-IRS element and between the $\ell$-th C-IRS element and the $u$-th RxV antenna, respectively; \textit{(ii)} $r_{\ell,k}^t$ and $r_{u,\ell}^r$ are propagation distances; \textit{(iii)} $\varrho_c(\cdot)$ is the C-IRS element pattern, function of the \textit{local} incidence/reflection angles $\boldsymbol{\vartheta}_{\ell,k}^{i,(c)} = (\theta_{\ell,k}^{i,(c)},\varphi_{\ell,k}^{i,(c)})$, $\boldsymbol{\vartheta}_{u,\ell}^{o,(c)}= (\theta_{\ell,k}^{o,(c)},\varphi_{\ell,k}^{o,(c)})$, identified by superscript$\,^{(c)}$; \textit{(iv)} $\varrho_t(\cdot)$, $\varrho_r(\cdot)$ are the single antenna patters at TxV and RxV respectively, function of the \textit{global} AoDs and AoAs $\boldsymbol{\vartheta}_{\ell,k}^t = (\theta_{\ell,k}^{t},\varphi_{\ell,k}^{t})$ and $\boldsymbol{\vartheta}_{u,\ell}^r= (\theta_{\ell,k}^{r},\varphi_{\ell,k}^{r})$. Notice that, for any metasurface profile, local incidence/reflection angles can be expressed as function of the global ones $\boldsymbol{\vartheta}_{\ell,k}^{i}$ and $\boldsymbol{\vartheta}_{u,\ell}^{o}$.
We assume the same model for both the TxV, RxV and C-RIS/C-IRS element pattern, making reference to the widely employed model for reflectarrays \cite{9569465, RadPatt}, yielding
\begin{equation}\label{eq:elementPattern}
    \varrho(\boldsymbol{\vartheta}) = \sqrt{2(2q+1)}\, \cos^{q}\left[\frac{\pi}{2} - \arcsin\left(\cos\theta\sin\varphi\right)\right]
\end{equation}
where $\boldsymbol{\vartheta}$ denotes any of the angles in \eqref{eq:generalModel}. The reflection matrix $\boldsymbol{\Phi}$ at the C-IRS in \eqref{eq:channelDefinition} is diagonal with entries defined as
\begin{equation}\label{eq:reflectingmatrix}
    \boldsymbol{\Phi} = \text{diag}\left(\beta_{1} e^{j\Phi_{1}},...,\beta_{\ell} e^{j\Phi_{\ell}},...,\beta_{MN} e^{j\Phi_{MN}}\right) 
\end{equation}
where $\beta_{\ell}$ and $\Phi_{\ell}$ denote the amplitude and phase of the $\ell=mn$-th element reflection coefficient, respectively. The overall channel impulse response between the $k$-th TxV array element and the $u$-th RxV array element is
\begin{equation}\label{eq:focusingChannel_singlaPath}
    \left[\mathbf{H}\right]_{(u,k)} = \Gamma_d  e^{-j\frac{2\pi}{\lambda} r^d_{k,u}} +
    \sum_{\ell } \Gamma_{\ell,k,u} 
    e^{-j\left(\frac{2\pi}{\lambda}(r_{k,\ell}^t + r_{u,\ell}^r) + \Phi_\ell\right)}
\end{equation}
where $\Gamma_d$ and $\Gamma_{\ell,k,u}$ incorporate the path loss and the element's gain for both direct and C-RIS/C-IRS channels, respectively. 

The path amplitudes in \eqref{eq:channelModel} and \eqref{eq:generalModel} depend on the path loss $PL$:
\begin{equation}\label{eq:complexGain}
    \alpha = \sqrt{\frac{1}{PL}} e^{j\xi}
\end{equation}
where the phase $\xi$ accounts for additional effects (e.g., Doppler shift) and it is assumed as uniformly distributed, i.e., $\xi\sim\mathcal{U}[0, 2\pi)$, independent across different paths. The propagation loss for the direct link with amplitude $\alpha_d$ is defined (in dB) as \cite{3GPPTR37885}
\begin{equation} \label{eq:PL0}
   PL_d = \underbrace{32.4 + 20\log_{10}(r_d) + 20\log_{10}(f)}_{\mu_{\text{LoS}}} + A_b + \chi
\end{equation}
where $r_d = \|\mathbf{p}_r-\mathbf{p}_t\|_2$ is the distance between the TxV and RxV, $f$ is the carrier frequency (in GHz), $\chi \sim \mathcal{N}(0, \sigma_{sh}^2)$ represents the lognormal distributed shadowing component, and $A_b \sim \mathcal{N}(\mu_b, \sigma^2_b)$ accounts for an additional attenuation due to blockage from $b$ vehicles simultaneously \cite{R1-1807672, 3GPPTR37885}. Assuming independence between the shadowing component and the blockage component, we can write the path-loss as
\begin{equation}\label{eq:tot_pl}
    PL_d \sim
    \mathcal{N}\left(\underbrace{\mu_{\text{LoS}}+\mu_b}_{\mu_{PL_b}}, \underbrace{\sigma_{sh}^2 + \sigma^2_b}_{\sigma_{PL_b}^2}\right)
\end{equation}
where $\mu_{\text{LoS}}$ is the deterministic term in \eqref{eq:PL0}. The model for the amplitudes of the  direct paths through the C-IRS $\alpha_{\ell,k}$, $\alpha_{u,\ell}$ in \eqref{eq:generalModel} is defined based on \cite{Gustafson6691924, tang2020wireless, Sanguinetti9300189}.

\section{Cylindrical RIS/IRS}\label{sect:reflection_conformal}

Here, we describe the generalized Snell reflection's law for an arbitrarily shaped surface, deriving the phase pattern to be applied to a cylindrical C-RIS that can reflect in arbitrary directions. To avoid the explicit knowledge of incident/reflection angles, we propose a novel angle-independent cylindrical C-IRS design, providing insights on manufacturing possibilities. 

\subsection{Generalized Reflection's Law}\label{subsect:geenralized_reflection}

Let us consider a plane wave impinging on a 3D reflecting surface $y = f(x,z)$ from an arbitrary direction defined by wavevector
\begin{equation}\label{eq:inWave}
\begin{split}
     \mathbf{k}&=-\frac{2\pi}{\lambda}\left[\sin\varphi_i\cos\theta_i,\,\sin\varphi_i\sin\theta_i,\,\cos\varphi_i\right]^\mathrm{T},
\end{split}
\end{equation}
where we assume the far-field condition, i.e., the incident wavefront on the C-RIS/C-IRS is planar. According to the channel model in Section \ref{subsect:channel_model}, this implies $\boldsymbol{\vartheta}^{t}_{\ell,k} \approx\boldsymbol{\vartheta}_{t} $, $\boldsymbol{\vartheta}^{r}_{u,\ell} \approx\boldsymbol{\vartheta}_{r} $, $\boldsymbol{\vartheta}^{i}_{\ell,k} \approx\boldsymbol{\vartheta}_{i} $, $\boldsymbol{\vartheta}^{o}_{u,\ell} \approx\boldsymbol{\vartheta}_{o} $.
The reflected wavevector, which can be derived using Snell's law \cite{Luneburg,doi:10.1126/science.1210713}, is defined as
\begin{equation}\label{eq:outWave}
\begin{split}
     \overline{\mathbf{k}}&=\frac{2\pi}{\lambda}\left[\sin\varphi_o\cos\theta_o,\,\sin\varphi_o\sin\theta_o,\,\cos\varphi_o\right]^\mathrm{T}.
\end{split}
\end{equation}
Notice that, by considering a reflecting surface, the refracted wave is negligible. Assuming there is a phase variation $\Phi(x,y,z)$ defined in the neighborhood of $y=f(x,z)$ (as it should admit the gradient in $(x, f(x,z), z)$),  
the relation between $\mathbf{k}$ and $\overline{\mathbf{k}}$ is determined by the generalized Snell's law \cite{Gutierrez:17, PhysRevApplied.9.034021}, that in vector form is given by
\begin{equation}\label{eq:gsl}
    \overline{\mathbf{k}} - \mathbf{k} = \nabla \Phi - k \mathbf{u}
\end{equation}
where $k\in \mathbb{R}$ and $\mathbf{u}$ is the unit vector normal to the surface, defined as
\begin{equation}\label{eq:normal}
    \mathbf{u} = \frac{\left[-\pdv{x}{f} , 1, -\pdv{z}{f} \right]^\mathrm{T}}{\sqrt{1 + \|\nabla f\|^2}}
\end{equation}
Based on the desired operation, to generate a specular or anomalous reflection, one can compute the corresponding phase profile using \eqref{eq:gsl}. The phase gradient $\nabla \Phi$ is tangential to the surface, therefore from \eqref{eq:gsl} and \eqref{eq:normal}, we obtain
\begin{equation}\label{eq:generalPhaseGradient}
    \nabla \Phi = \overline{\mathbf{k}} - \mathbf{k}
    - \delta \left[-\pdv{x}{f} , 1, -\pdv{z}{f} \right]^\mathrm{T}
\end{equation}
where
\begin{equation}
    \delta = \frac{(\overline{\mathbf{k}} - \mathbf{k})^\mathrm{T} \left[-\pdv{x}{f} , 1, -\pdv{z}{f} \right]^\mathrm{T}}{1 + \|\nabla f\|^2}.   
\end{equation}
Although \eqref{eq:generalPhaseGradient} contains three differential equations, only two equations are mutually independent, i.e., $\Phi(x,y,z) = \Phi(x, f(x,z), z)$. Therefore, the phase increment along any tangential direction on the surface can be expressed as
\begin{equation}\label{eq:phaseIncrement}
    \mathrm{d}\Phi = \pdv{x}{\Phi} \mathrm{d}x + \pdv{y}{\Phi} \mathrm{d}y + \pdv{z}{\Phi} \mathrm{d}z
    =\left(\pdv{x}{\Phi} + \pdv{y}{\Phi}\pdv{x}{y}\right)\mathrm{d}x + \left(\pdv{z}{\Phi} + \pdv{z}{\Phi}\pdv{z}{y}\right)\mathrm{d}z.
\end{equation}
By substituting \eqref{eq:generalPhaseGradient} into \eqref{eq:phaseIncrement}, we obtain
\begin{equation}\label{eq:phaseIncrementSolved}
\begin{split}
    \mathrm{d}\Phi = \left( \overline{k}_x - k_x + (\overline{k}_y - k_y) \pdv{x}{f}\right) \mathrm{d}x +
    \left(\overline{k}_z - k_z + (\overline{k}_y - k_y) \pdv{z}{f}\right) \mathrm{d}z
\end{split}
\end{equation}
and by integration
\begin{equation}\label{eq:generalPhase}
    \Phi(x, y, z) = (\overline{k}_x - k_x) x + (\overline{k}_y - k_y) y + (\overline{k}_z - k_z) z.
\end{equation}
By plugging an arbitrary 3D position of the $(m,n)$-th C-RIS/C-IRS element into \eqref{eq:generalPhase}, namely setting $x = x_{m,n}$, $y = y_{m,n}$ and $z = z_{m,n}$, we obtain the optimal phase configuration as
\begin{multline}\label{eq:phaseConfiguration}
    \Phi_{m,n} = - \frac{2\pi}{\lambda}\left[ x_{m,n}\, (\cos \theta_o \sin \varphi_o + \cos \theta_i \sin \varphi_i) + 
    y_{m,n} \,(\sin \theta_o \sin \varphi_o + \sin \theta_i \sin \varphi_i) + \right.\\
    \left. +z_{m,n}\, (\cos \varphi_o + \cos \varphi_i)\right].
\end{multline}
Herein, we set $x_{m,n}$, $y_{m,n}$ and $z_{m,n}$ as in Section \ref{sect:system_model}, to represent a cylindrical surface of radius $R$ approximating the shape of the car's body with a quadratic profile in the vertical direction and no curvature in the horizontal one. It is worth noticing that \eqref{eq:phaseConfiguration} can represent any surface profile and can be used to generate specular or anomalous reflections knowing the desired incident $(\theta_i, \varphi_i)$ and reflected $(\theta_o, \varphi_o)$ angles. When $R\rightarrow\infty$, $x_{m,n}\rightarrow 0$, $z_{m,n}\rightarrow d_m(m-1)$, \eqref{eq:phaseConfiguration} yields the phase configuration for planar RIS/IRS \cite{direnzo2021surveyRIS}
\begin{equation}\label{eq:farfield_phase}
    \Phi^\infty_{m,n} = -\frac{2\pi}{\lambda} \left[d_n(n-1) (\sin \theta_o \sin \varphi_o + \sin \theta_i \sin \varphi_i) + 
   d_m (m-1) (\cos \varphi_o + \cos \varphi_i) \right].   
\end{equation}
Therefore, the phase configuration \eqref{eq:phaseConfiguration} can be regarded as generalization of the planar one.

\subsection{Pre-configured C-IRS Design}

Dynamic configuration of the C-RIS requires accurate a-priori information of the incident and reflection angles $(\theta_i,\varphi_i)$ and $(\theta_o,\varphi_o)$, which can be obtained through signaling between the TxV, RxV, and C-RIS. However, in a V2V scenario, these angles can rapidly change, unbearably increasing the signaling overhead. Hence, we herein propose a novel phase design for a pre-configured, fully passive cylindrical-shaped C-IRS, that can be a local approximation on a portion of car lateral doors. The proposed design does not require the knowledge of time-varying incidence and reflection angles across the C-IRS elements, easing the practical deployment in vehicular scenarios.

We approach the C-IRS design by splitting the whole 3D problem into two distinct 2D incidence/reflection problems, considering an incident plane wave lying in the $x-z$ elevation plane (across the cylindrical direction) and another incident wave in the $x-y$ azimuth plane (along the cylindrical direction), respectively characterized by wavevectors
\begin{align}
    \mathbf{k}_{xz}&=-\frac{2\pi}{\lambda}\left[\sin\varphi_i,\,0,\,\cos\varphi_i\right]^\mathrm{T}\label{eq:Kxz}\\
    \mathbf{k}_{xy}&=-\frac{2\pi}{\lambda}\left[\cos\theta_i,\,\sin\theta_i,\,0\right]^\mathrm{T}\label{eq:Kxy}
\end{align}
obtained from \eqref{eq:inWave} by setting, respectively, $\theta_i=0$ (elevation plane) and $\varphi_i=\pi/2$ (azimuth plane).
Notice that we are not aiming at decomposing the 3D wavevector, thus $\mathbf{k}\neq\mathbf{k}_{xy} + \mathbf{k}_{xz}$, but rather we separate the plane wave of incidence to highlight and discuss the specific phenomena. 

\subsubsection{\textbf{Incident wave in the $x-z$ plane (elevation plane)}}

\begin{figure}[!b]
    \centering
    \includegraphics[width=0.5\columnwidth]{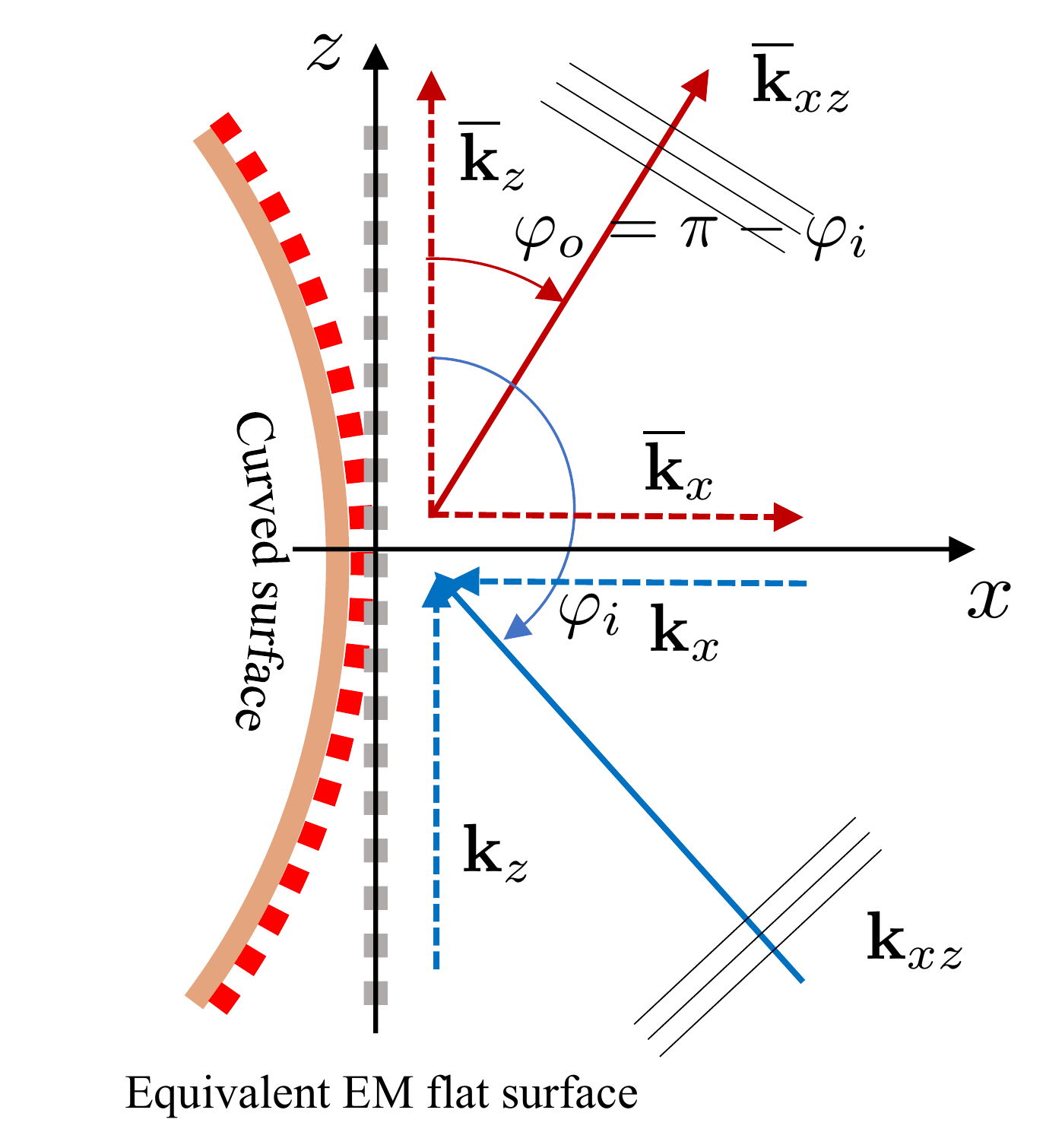}
    \caption{Plane wave incidence on the $x-z$ plane and specular reflection. The phase configuration across the C-IRS is designed such that to make the curved surface to behave like a EM flat surface}
    \label{fig:xzplane}
\end{figure}
\begin{figure}[b!]
    \centering
    \includegraphics[width=0.7\columnwidth]{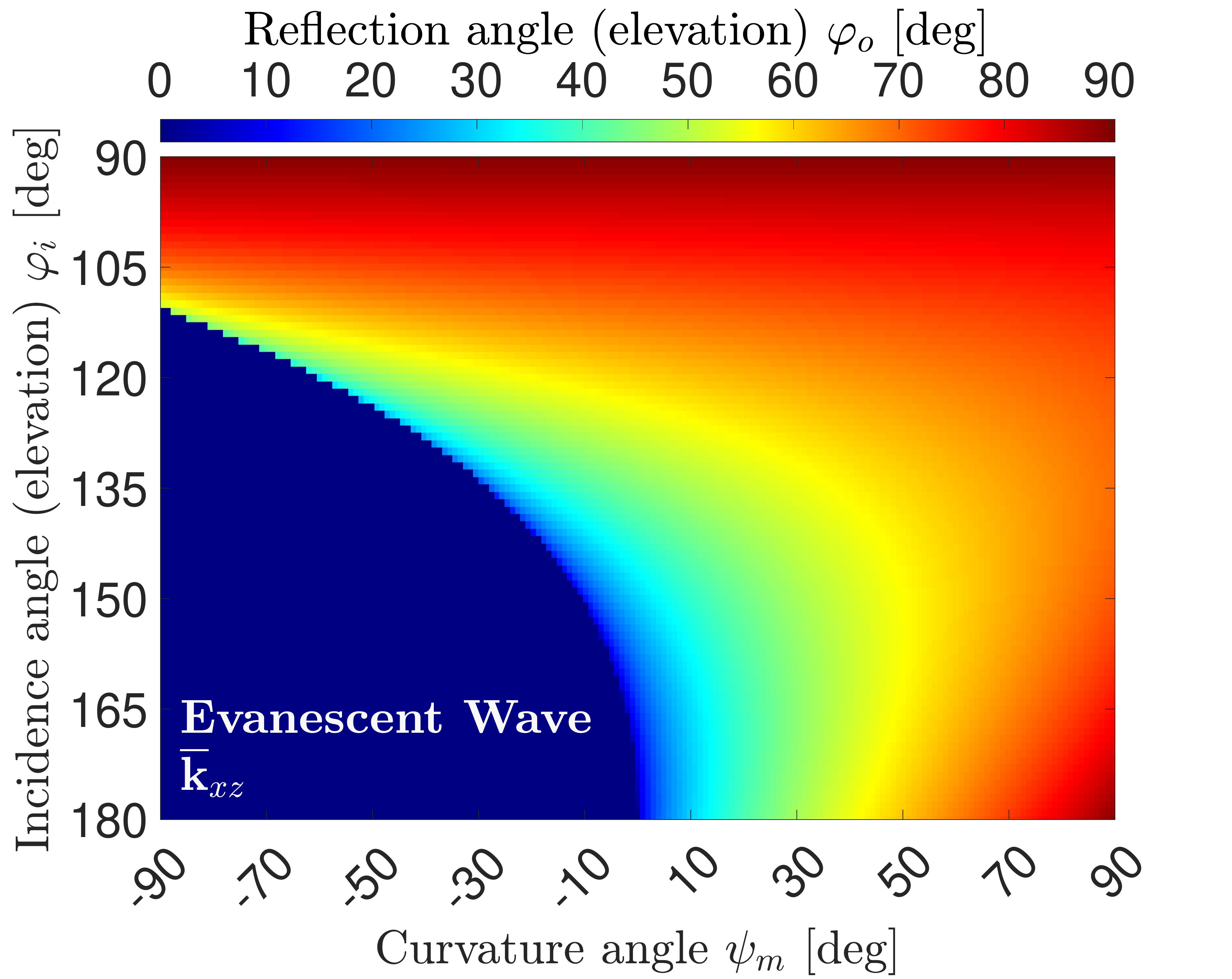}
    \caption{Reflection angle $\varphi_o$ in \eqref{eq:phi_out} as function of the incidence angle $\varphi_i$ and of the specific incidence point on the C-IRS (defined by $\psi_m$)}
    \label{fig:PhiOut}
\end{figure}

Fig. \ref{fig:xzplane} depicts an incident planar wave on the conformal direction of the C-IRS.  Let us assume the surface is made of a fully reflecting material (e.g., a metal); thus, there is no penetration. For a bare cylindrical surface, all the incident rays are reflected toward different angles, following the Snell's law applied to the local elevation angle $\varphi_i^{(c)}$ at each point on the surface. Therefore, the wave's energy is scattered in space (neglecting border effects), and only a limited portion can be captured by a receiver placed at AoD $\varphi_o$. 
Now assume that $M$ configurable elements cover the cylindrical surface. The optimal phase profile, given incidence and reflection elevation angles $\varphi_i$ and $\varphi_o$, is 
\begin{equation}\label{eq:phase-x-z}
    \begin{split}
        \Phi_m = \frac{8\pi R}{\lambda} \left[\sin\left(\frac{\psi_m}{2}\right) \cos\left(\frac{\varphi_o-\varphi_i}{2}\right)\cos\left(\frac{\varphi_o+\varphi_i+\psi_m}{2}\right)\right]
    \end{split}
\end{equation}
The latter is derived from \eqref{eq:phaseConfiguration} by setting $\theta_i =\theta_o=0$. Again, the optimal phase profile requires the perfect knowledge (or at least an accurate estimation) of the incidence/reflection angles, as well as the position of C-IRS elements. While the elements' relative position is fixed and can be taken into account during the manufacturing process, $\varphi_i$ and $\varphi_o$ are not a-priori known and cannot be included in the C-IRS design.    

\begin{figure}[t!]
    \centering
    \subfloat[][$R=2$ m]{\includegraphics[width=0.5\columnwidth]{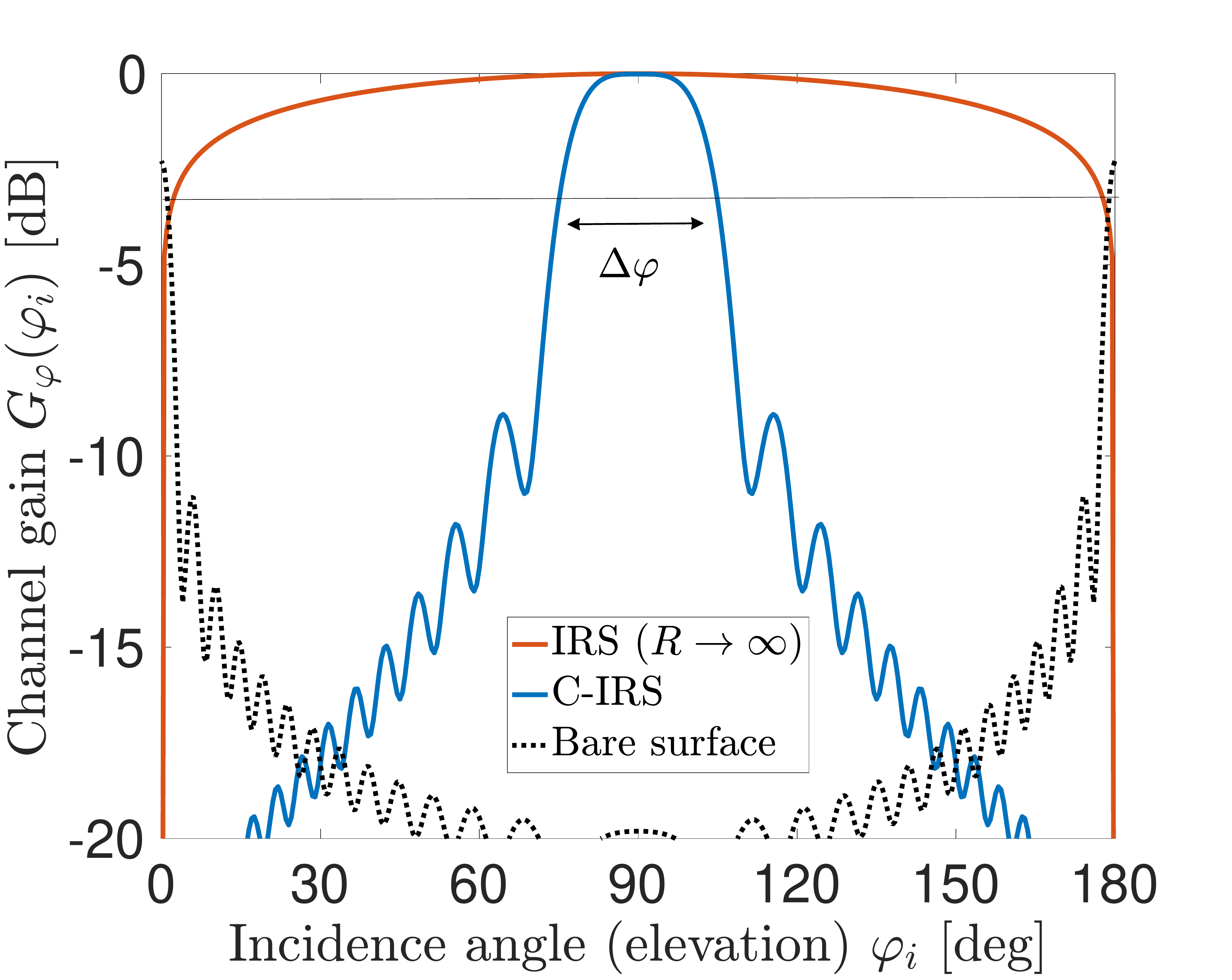}\label{subfig:GvsPhi_2m}} 
    \subfloat[][$R=8$ m]{\includegraphics[width=0.5\columnwidth]{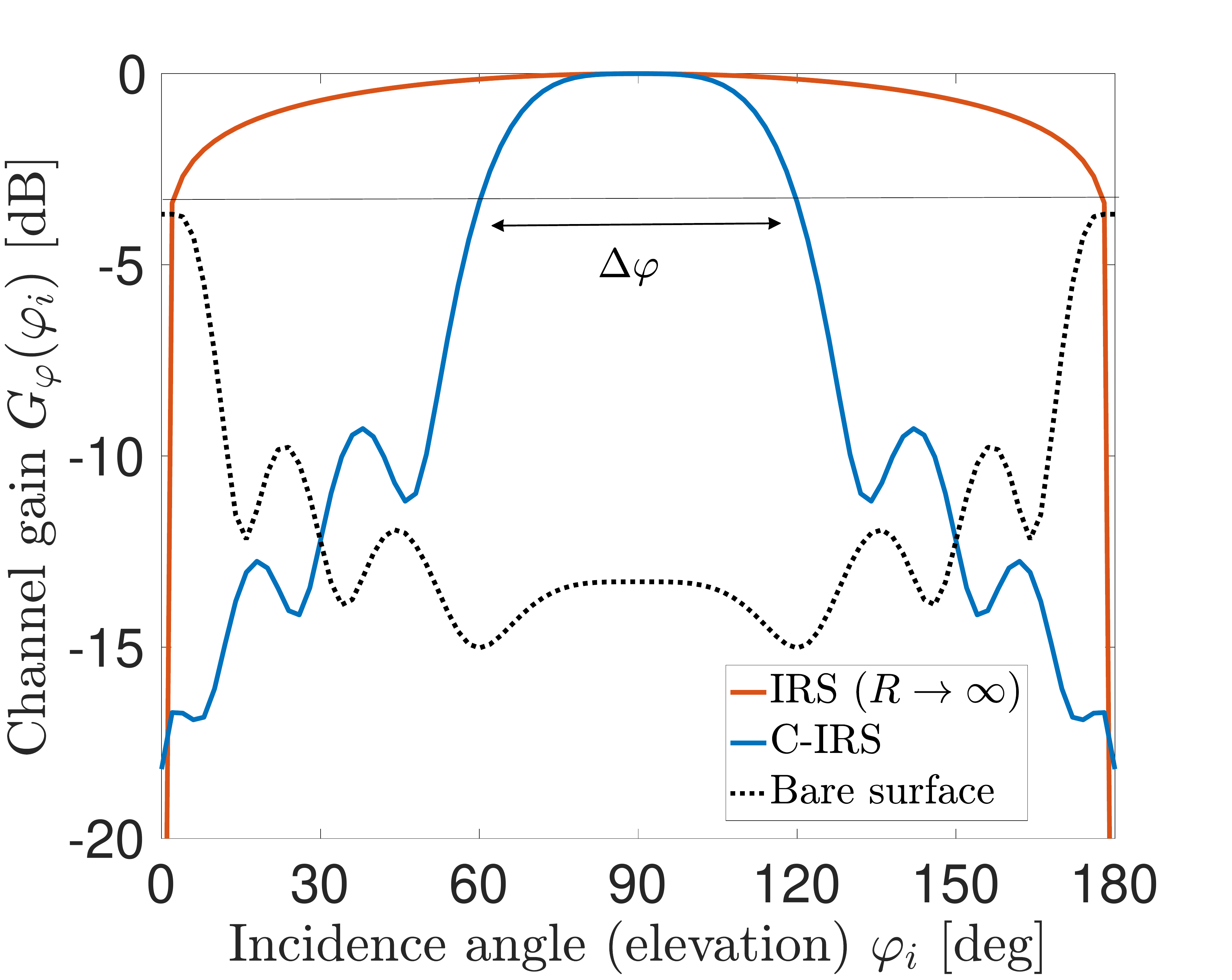}\label{subfig:GvsPhi_8m}} 
    \caption{Channel gain on the elevation plane for a C-IRS compared to a flat IRS ($R\rightarrow\infty$) and a bare cylindrical surface (i.e., without phase compensation) for (\ref{subfig:GvsPhi_2m}) $R=2$ m and (\ref{subfig:GvsPhi_8m}) $R=8$ m. }
    \label{fig:GvsPhiPho}
\end{figure}

To obviate this issue, the phase pattern across the C-IRS can be designed to reflect back only the $x$ component of wavevector $\mathbf{k}_{xz}$, perpendicular to the desired flat surface (Fig. \ref{fig:xzplane}). Therefore, let us consider the decomposition $\mathbf{k}_{xz} = \mathbf{k}_{x} + \mathbf{k}_{z}$, where 
\begin{align}
    \mathbf{k}_{x} = -\frac{2 \pi}{\lambda}\left[\sin\varphi_i,\,0,\,0\right]^\mathrm{T}, \qquad
    \mathbf{k}_{z} = -\frac{2 \pi}{\lambda}\left[0,\,0,\,\cos\varphi_i\right]^\mathrm{T},
\end{align}
denote, respectively, the perpendicular and parallel components of the incident wavevector $\mathbf{k}_{xz}$ \eqref{eq:Kxz}. The goal is to mimic a perfectly flat surface enabling  a strong specular reflection, namely back reflecting $\mathbf{k}_x$ while preserving $\mathbf{k}_{z}$ at each point on the C-IRS, i.e.,
\begin{align}
    \overline{\mathbf{k}}_x = -\mathbf{k}_x,\qquad
    \overline{\mathbf{k}}_z = \mathbf{k}_z.
\end{align}
Setting $\varphi_i = \varphi_o = \pi/2$ in \eqref{eq:phase-x-z} yields the following phase profile across the C-IRS elements 
\begin{equation}\label{eq:phasePerpendicular}
    \Phi^\perp_m = -\frac{4 \pi R}{\lambda} \left(\cos\psi_m -1 \right),
\end{equation}
where superscript $\,^\perp$ indicates the proposed phase configuration. Phase \eqref{eq:phasePerpendicular} depends \textit{only} on the shape of the C-IRS along the conformal direction, i.e., on curvature radius $R$ and angular position $\psi_m$. By configuring the phase pattern based on \eqref{eq:phasePerpendicular}, we ensure that only the $x$ component of the wavevector, $\mathbf{k}_x$, is specularly reflected, while the response of the C-IRS to the $z$ component $\mathbf{k}_{z}$ depends on both the specific incidence point onto the C-IRS, i.e., $\psi_m$, and the incidence elevation angle $\varphi_i$. By substituting equation \eqref{eq:phasePerpendicular} into equation \eqref{eq:phase-x-z} we can derive the elevation angle $\varphi_o$ of the reflected wave $\overline{\mathbf{k}}_{xz}$ as
\begin{equation}\label{eq:phi_out}
    \varphi_o = \arccos\left[-2\sin\left(\frac{\psi_m}{2}\right) \hspace{-0.1cm}- \cos\left(\varphi_i + \frac{\psi_m}{2}\right)\right] - \frac{\psi_m}{2}
\end{equation}
Based on \eqref{eq:phi_out}, the reflected wavevector $\overline{\mathbf{k}}_{xz}$ can experience different propagation effects depending on the incidence angle $\varphi_i$ and $m$-th element position (defined by curvature angle $\psi_m$), as depicted in Fig. \ref{fig:PhiOut}. In particular, when $\varphi_i = \pi/2$ (incident wave perfectly perpendicular to the C-IRS, $\mathbf{k}_{xz}=\mathbf{k}_{x}$) we obtain that $\varphi_o = \pi/2$ $\forall m$. Thus, every C-IRS element is illuminated by the impinging wave and we have a perfect specular reflection. When $\varphi_i \in \{0,\pi\}$ (incident wave perfectly parallel to the $z$ axis, $\mathbf{k}_{xz}=\mathbf{k}_{z}$), we observe that $\varphi_o =\{\pi,0\}$ only for $\psi_m = 0$, while $\varphi_o = \pi/2$ for $\psi_m = \pi/2$: only the edge element of the C-IRS provides the desired specular reflection. For all the intermediate angles $0\leq\varphi_i\leq \pi/2$ (or equivalently $\pi/2\leq\varphi_i\leq \pi$), only a portion of the C-IRS elements contribute to the reflection, while for all the others, the reflected wave $\overline{\mathbf{k}}_{xz}$ is evanescent and thus attenuated. The latter condition is achieved for:
\begin{equation}
    \left|-2\sin\left(\frac{\psi_m}{2}\right) - \cos\left(\varphi_i + \frac{\psi_m}{2}\right)\right| > 1.
\end{equation}
The impact of the phase configuration in \eqref{eq:phasePerpendicular} can be assessed by the normalized channel gain, varying the incidence angle $\varphi_i$ and evaluated in $\varphi_o=\pi-\varphi_i$ (specular reflection), with and without the C-IRS (bare surface), for $R=2$ m (Fig. \ref{subfig:GvsPhi_2m}) and $R=8$ m (Fig. \ref{subfig:GvsPhi_8m}).
The operating frequency is $f=28$ GHz (wavelength $\lambda=1.07$ cm). The normalized channel gain is defined as
\begin{equation}\label{eq:normalized_channel_gain_elev}
    G_\varphi(\varphi_i) = \mathrm{tr}\left(\mathbf{H}_{cr} \boldsymbol{\Phi} \mathbf{H}_{tc} \mathbf{H}^\mathrm{H}_{tc} \boldsymbol{\Phi}^\mathrm{H} \mathbf{H}^\mathrm{H}_{cr}\bigg\lvert_{\substack{\theta_i=\theta_o=0}}\right)
\end{equation}
where $\mathbf{H}_{tc}$ and $\mathbf{H}_{cr}$ are from \eqref{eq:generalModel} and normalized such that $\|\mathbf{H}_{tc}\|_F = \|\mathbf{H}_{cr}\|_F = \|\mathbf{\Phi}\|_F = 1$. The channel gain $G_\varphi(\varphi_i)$ does not consider the path-loss, thus it is inherently independent on distances to highlight relative channel gains. We evaluate $G_\varphi(\varphi_i)$ for $A=1$ m$^2$. The result shows that, in the considered setting, the C-IRS performance (blue solid line) practically matches the flat IRS one (red solid line) over an angular interval $\Delta\varphi$ of $\approx 20$ deg (defined at $-3$ dB from the peak) for $R=2$ m and $\approx 40$ deg for $R=8$ m. Thus, the C-IRS will optimally reflect the incident wave for $\varphi_i\in[\pi/2-(\Delta\varphi/2),\pi/2+(\Delta\varphi/2)]$. Most important, the usage of a C-IRS provides a relative channel gain in excess of $15$ dB compared to a bare cylindrical surface (black dotted line).
positive slopes and $\psi_c \in [-90^{\circ},0]$ for negative slopes, this fact is automatically taken into account.

\begin{figure}[t!]
    \centering
    \subfloat[][$R=2$ m]{\includegraphics[width=0.45\columnwidth]{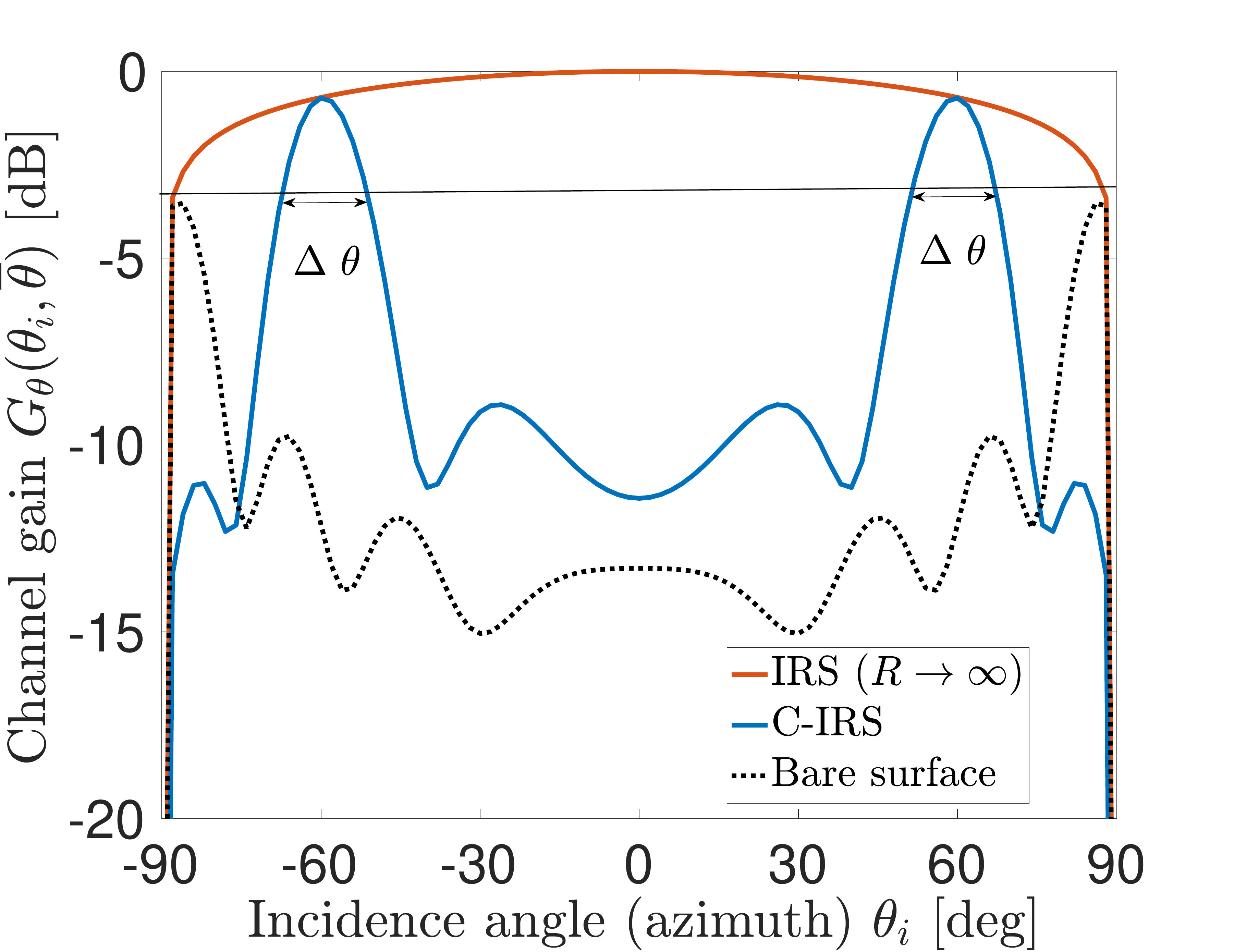}\label{subfig:GvsThi_2m}}
    \subfloat[][$R=8$ m]{\includegraphics[width=0.45\columnwidth]{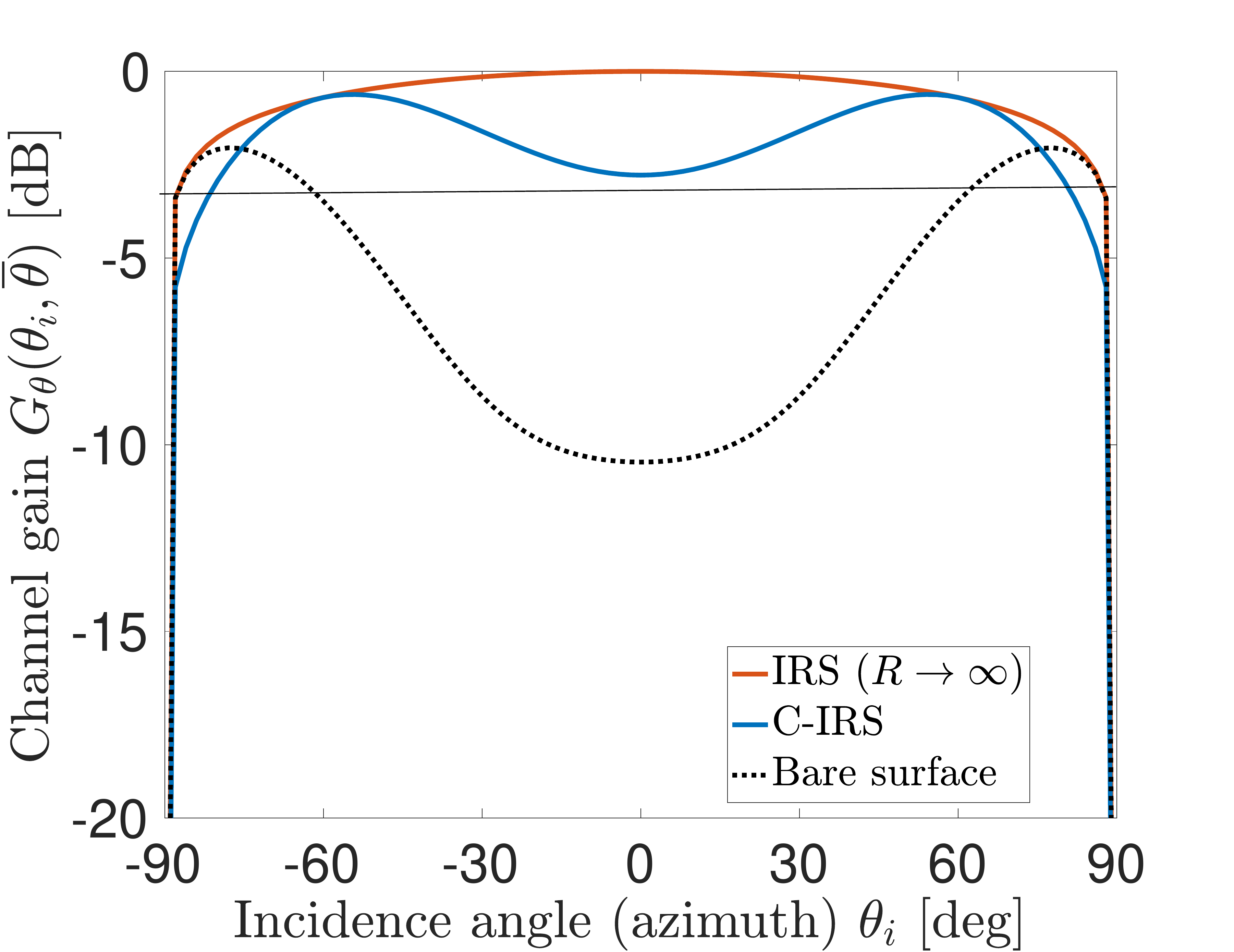}\label{subfig:GvsThi_8m}}
    \caption{Channel gain for C-IRS compared to flat IRS and a bare conformal surface (without phase compensation) for (\ref{subfig:GvsPhi_2m}) $R=2$ m and (\ref{subfig:GvsPhi_8m}) $R=8$ m . }
    \label{fig:GvsThi}
\end{figure}

\subsubsection{\textbf{Incident wave in the $x-y$ plane (azimuth plane)}}
The incidence on the $x-y$ plane implies $\varphi_i = \pi/2$ and it is characterized by the wavevector defined in \eqref{eq:Kxy}. The optimal phase profile, derived from \eqref{eq:phaseConfiguration}, is 
\begin{equation}\label{eq:phase-x-y}
    \begin{split}
        \Phi_{m,n} = -\frac{4\pi}{\lambda} \cos\left(\frac{\theta_o - \theta_i}{2}\right) \left[R (\cos\psi_m -1)  \cos\left(\frac{\theta_o + \theta_i}{2}\right) + d_n(n-1) \sin\left(\frac{\theta_o + \theta_i}{2}\right) \right].
    \end{split}
\end{equation}
It depends on the curvature angle $\psi_m$ as well as on both the incidence and reflection azimuth angles $\theta_i$ and $\theta_o$, and, differently from \eqref{eq:phase-x-z}, it requires a phase gradient along the cylindrical direction, thus a 2D phase configuration.
To achieve a specular reflection on the azimuth plane, we set $\theta = \theta_i = -\theta_o$ in \eqref{eq:phase-x-y}, obtaining the 1D phase pattern:
\begin{equation}\label{eq:xy-specular}
    \Phi^\perp_m = -\frac{4 \pi R}{\lambda} \left(\cos\psi_m -1 \right) \cos\theta,
\end{equation}
that is the same phase profile of \eqref{eq:phase-x-z} except for an additional multiplicative term depending on $\theta$. It follows that, as a consequence of the curvature along $x$, it is again necessary to know the azimuth angle of incidence $\theta_i$ (or reflection $\theta_o$) to obtain a specular reflection. However, the goal of the proposed solution is to produce a fully-passive C-IRS; thus, non-reconfigurable. Therefore, it is necessary to design the C-IRS with an azimuth angle $\overline{\theta}\neq 0$ that can serve most of the V2V links. By fixing $\overline{\theta}$, the C-IRS will optimally reflect when the incidence (or reflection) is $\theta_i \in [\overline{\theta} - (\Delta \theta/2), \overline{\theta}+ (\Delta \theta/2)]$, where $\Delta \theta$ defines the width of an angular interval in azimuth where the specular reflection is guaranteed, i.e., the energy of the reflected wave in the specular direction is maximum.

The angular bandwidth $\Delta \theta$ can be empirically derived from the normalized channel gain on the azimuth plane as function of $\theta_i$ for $\theta_i=-\theta_o$ (specular reflection):
\begin{equation}\label{eq:normalized_channel_gain_az}
    G_\theta(\theta_i,\overline{\theta}) = \mathrm{tr}\left(\mathbf{H}_{cr} \boldsymbol{\Phi} \mathbf{H}_{tc} \mathbf{H}^\mathrm{H}_{tc} \boldsymbol{\Phi}^\mathrm{H} \mathbf{H}^\mathrm{H}_{cr}\bigg\lvert_{\substack{\varphi_i=\varphi_o=\frac{\pi}{2}}}\right).
\end{equation}
Fig. \ref{fig:GvsThi} shows $G_\theta(\theta_i,\overline{\theta})$ varying $\theta_i$ for $\overline{\theta}=\pi/3$, $f_0=28$ GHz, $R=2$ m (Fig. \ref{subfig:GvsThi_2m}) and $R=8$ m (Fig. \ref{subfig:GvsThi_8m}), fixing the area of the C-IRS to $A=1$ m$^2$. As expected, the more the cylindrical surface tends to a flat surface ($R\rightarrow\infty$), the larger is the angular bandwidth of specular reflection $\Delta \theta$. In particular, for $R=2$ m (highly curved surface), the channel gain exhibits a an azimuth selectivity, namely $\Delta\theta \approx 15$ deg, while $\Delta\theta \approx 90$ deg (no angular selectivity) for $R=8$ m (slightly curved surface). 

\section{C-IRS for vehicular applications}\label{subsect:CIRS_vehicular}

\begin{figure}[t!]
    \centering
    \subfloat[][]{\includegraphics[width=0.5\columnwidth]{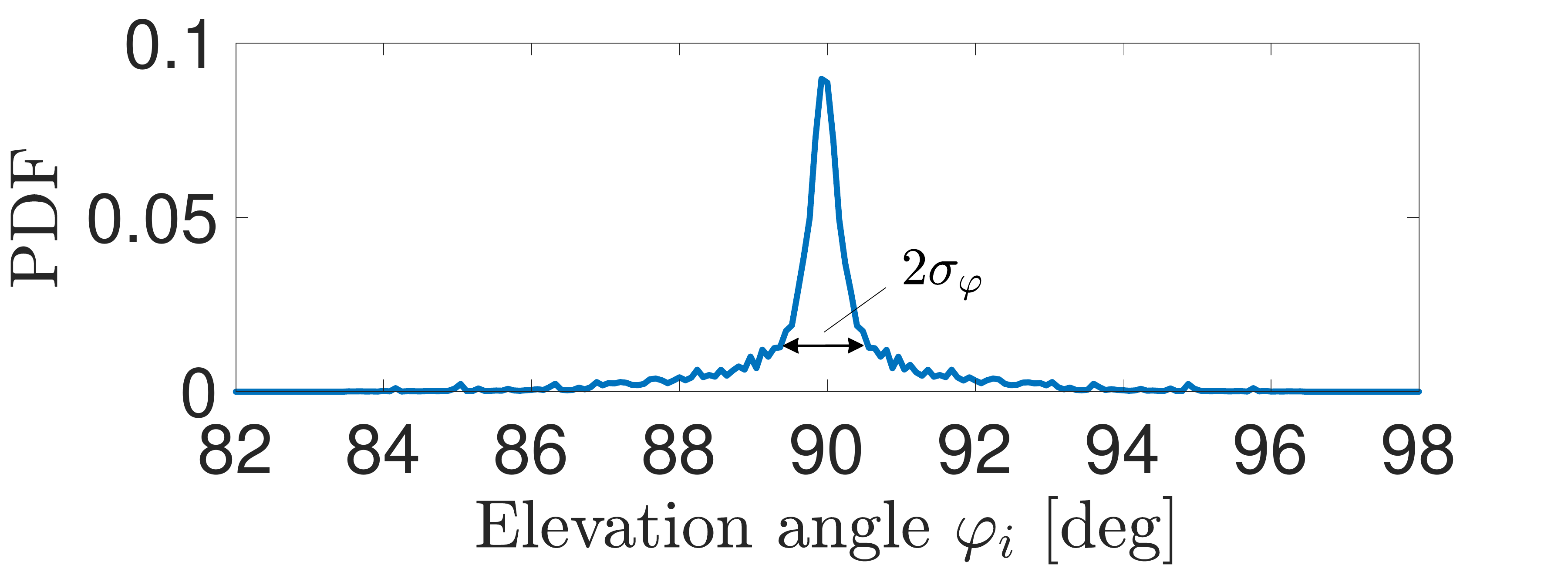}\label{subfig:ELpdf}}
    \subfloat[][]{\includegraphics[width=0.5\columnwidth]{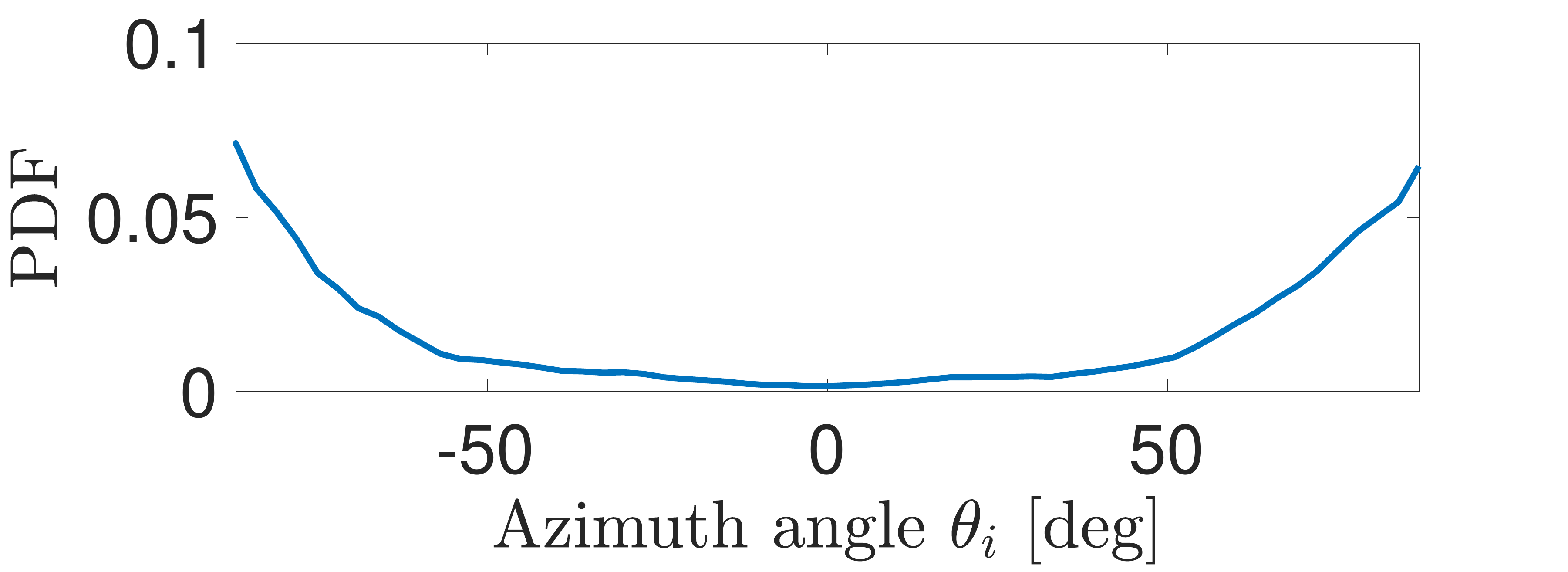}\label{subfig:AZpdf}}
    \caption{Empirical PDF of (\ref{subfig:ELpdf}) elevation angles $\varphi$ and (\ref{subfig:AZpdf}) azimuth angles in a V2V highway scenario (incidence/reflection)}
    \label{fig:EL&AZPdf}
\end{figure}

The operative conditions for the C-IRS in vehicular scenarios depend on the statistical distribution of elevation and azimuth angles of incidence onto the C-IRS, $\varphi_i$ and $\theta_i$. The empirical probability density function (PDF) of the impinging angle $\varphi_i$ on a C-IRS-equipped vehicle's doors when considering a highway V2V scenario is shown in Figs. \ref{subfig:ELpdf} and \ref{subfig:AZpdf}. The PDF is obtained through extensive simulation based on SUMO \cite{SUMO2018} for different vehicle types, e.g., passenger cars, trucks, buses, and these resemble those in \cite{MIZMIZI2022100402, 9473772}.
Since vehicles have similar heights compared to typical V2V distances (ranging from few to tens of meters), the distribution of elevation angles can be approximated by a Gaussian PDF where with a standard deviation of $\sigma_\varphi\approx 1.5$ deg, centered around $\varphi_i=\pi/2$. According to Fig. \ref{fig:PhiOut}, for $\varphi_i\in[\pi/2-\sigma_\varphi,\pi/2+\sigma_\varphi]$ the reflection angle $\varphi_o$ is within the same angular interval of $\varphi_i$, thus $\varphi_o\in[\pi/2-\sigma_\varphi,\pi/2+\sigma_\varphi]$. In this angular range, the TxV-C-IRS-RxV channel gain $G_\varphi(\varphi_i)$ does not show an appreciable reduction compared to the ideal case at $\varphi_i=\varphi_o=\pi/2$, as reported in Fig. \ref{fig:GvsPhiPho}. Therefore, $\Delta\varphi \gg 2\sigma_\varphi$, i.e., the C-IRS guarantees the specular reflection for all the angles within the typical V2V occurrence, regardless of the curvature of the car door $R$.

Differently from elevation, the PDF of incidence azimuth angle $\theta_i$ is wider and monotonically increases for $|\theta_i|\rightarrow\pi/2$, as shown in Fig. \ref{subfig:AZpdf}. The azimuth variability is a consequence of the random positions in space of TxV, RxV and of the relaying vehicle (varying of tens of meters), that dominates over the height differences ruling the PDF of $\varphi_i$ (in the order of tens of centimeters to few meters). In any case, the angular interval for which the C-IRS provides a specular azimuth reflection explicitly depends on parameter $\overline{\theta}$, as shown in Fig. \ref{fig:GvsThi}, and in the worst possible case ($R=2$ m, highly curved car door) it amounts to $\Delta\theta\approx 15$ deg. Therefore, $\overline{\theta}$ can be pre-configured to make the C-IRS work around the dominant region of the PDF, i.e., for $|\theta_i|\geq \pi/3$, thus $\pi/3\leq\overline{\theta}\leq\pi/2$, depending on the average TxV-RxV distance (Section
\ref{sect:numerical_results}).

\begin{figure}[!t]
    \centering
    \includegraphics[width=0.55\columnwidth]{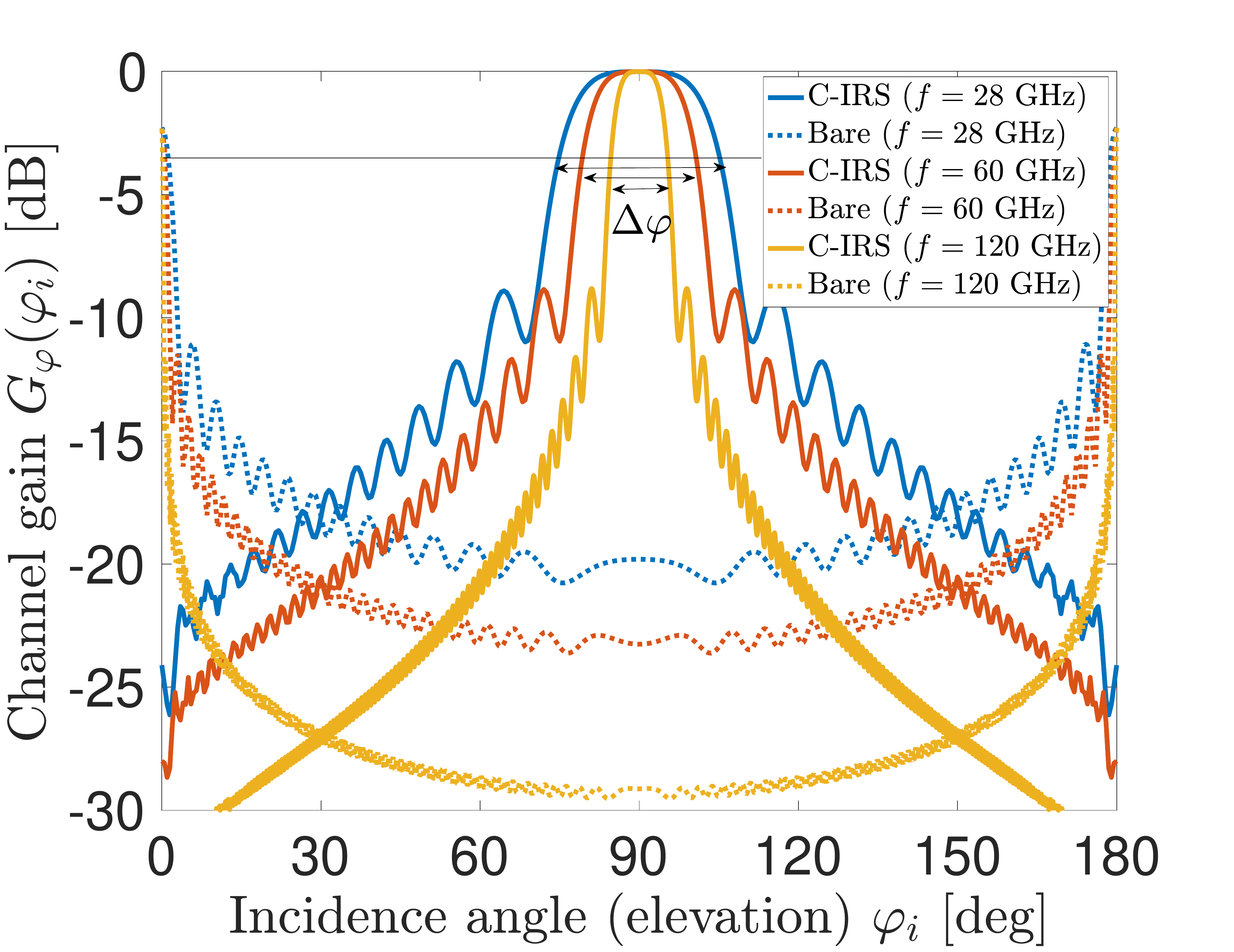}
    \caption{Normalized channel gain on the elevation plane varying the incidence angle $\theta_i$, for $\theta_o=-\theta_i$ (specular reflection) and different values of operating frequency.   }
    \label{fig:GvsPhivsf0}
\end{figure}

It is interesting to show the trend of the normalized channel gain, for instance on the elevation plane, $G_\varphi(\varphi_i)$ varying the operating frequency $f$, for a fixed curvature radius $R$ and the area of the C-IRS ($A=1$ m$^2$). While increasing the frequency (fixing the C-IRS physical area $A$) provides more gain in the specular direction, it also leads to a progressive reduction of the available angular interval $\Delta\varphi$ for specular reflection. The latter effect is illustrated in Fig. \ref{fig:GvsPhivsf0}, for $R=2$ m and $f=28,\,60,\,120$ GHz. The C-IRS phase configuration is given by \eqref{eq:phasePerpendicular}. The choice of the operating frequency $f$ is therefore ruled by a number of factors, among which the manufacturing possibilities play a major role, as illustrated in the following subsection.


\subsection{Realization of C-IRS}\label{subsect:CIRS_realization}

C-IRS design is accomplished by the general concept of the metasurface, i.e., a 2D array of periodic or quasi-periodic patterned scatterers (metallic or dielectric) causing anomalous reflection/transmission by an additional space-dependent phase shift introduced by the scatterers \cite{G1}. Well-established design criteria can be used for the practical implementation of C-IRS. Reflectarrays/transmitarray concept in which elements size and therefore spacings are of the order of the wavelength \cite{G2}, offer ''discrete'' beamforming capabilities since the phase pattern is sampled at the various elements positions. The ideal metasurface concept, in which elements and spacings are a small fraction of the wavelength, provides additional flexibility in the design since the phase pattern that can be potentially synthesized is almost continuous. This latter implementation is far more complex
and, as a further drawback, it can result in exceedingly high ohmic losses, so that in practice, the concept of reflectarray/transmitarray is by far the most used one and even more so when low-cost implementations are in order.  
Sophistications corresponding to multilayer metasurfaces or reflectarrays/transmitarrays have been proposed too \cite{G3}. Each tile in the metasurface can offer additional flexibility in the design, such as scattering to different directions, multi-frequency operation, or holographic-type wave processing \cite{G4}, \cite{G5}. 

The design of the master element (or master {\emph{elements}}) in the quasi-periodic patterned array is a key step in the design process since realization tolerances and, therefore, the final performance of the C-IRS are affected by its choice. Design specifications, such as bandwidth, type of polarization (linear or double), and efficiency, are determined almost exclusively by the master element/elements. Low-cost implementation is best addressed by metallic patterning on simple substrates, which allows for double-polarization capabilities, good efficiency, and robustness to fabrication tolerances at least up to mmWave frequencies. The absolute accuracy and repeatability are mainly dependent on the accuracy of the lithographic process.

C-IRS practical realization must account for several factors, such as deployment on curved/shaped surfaces with varying local curvature parameters, low profile, and low cost, with a view to massive production. Flexible polymer materials, such as polyethylene terephthalate (PET), PVC, and polyimide \cite{G6} \cite{G7}, offer versatility in terms of thickness and permittivity values and are suitable for outdoor applications because of their good thermal properties and robustness. PET has been previously used in antenna design for 5G operation \cite{G8} and is compatible with low-cost ink-jet printing and, its intrinsic losses are of the order of $10^{-2}$ in X-band, which is a reasonable value also for C-IRS. Polyimide substrates can adhere to metals grounds without the need for adhesive layers and, they can also be used as external insulating and passivating layers. Polyimide offers a good loss factor ($10^{-3}$ in C-band) and, it has been used for up to sub-THz frequencies with good performances (loss tangent  $10^{-2}$ at 1 THz). 
Metal reactive elements can be introduced, e.g., by screen printing or inkjet printing of silver nanoparticles. Inkjet can offer good resolution and is currently able to realize features of the order of $100 \mu$m and below with an accuracy that mainly depends on the silver nanoparticles size \cite{G10}. Silver provides extremely good conductivity, is less costly than gold, and, in contrast to copper, is almost unaffected by oxide formation. In the literature, high values of conductivity of inks embedding silver nanoparticles have been obtained even close to the bulk conductivity of silver ($\simeq 6.2\cdot 10^{-7}$ S/m). Nanoparticle size is a key aspect in the realization of small details and, current technology demonstrated the realization of very homogeneous particle size distributions with a mean value around $60$ nm \cite{G11}. 
Research on inkjet printing in the high-frequency application is nowadays quite advanced and, its use up to 80 GHz has been demonstrated in critical components such as narrowband filters. Details as small as 25 um have been claimed and, the main drawback is represented by the residual surface roughness of the ink, which is also strongly dependent on the surface roughness of the substrate used \cite{G11}. In the frame of C-IRS realization, this does not represent an issue, since the quality factor of the resonant element in the C-IRS is not a critical parameter in the design process. Some edge effects in the printing process can contribute to deviations from the ideal behavior of patterned metal surfaces. In contrast to etching techniques, in which the sides of the patterned metallizations can have different vertical shapes according to the thickness of the copper and etching parameters, inkjet printing shows a repeatable \emph{undercutting} like behavior at vertical edges. Such effect can be compensated in the design once the printing process parameters have been fixed in the final engineering of the surface. Inkjet printing also offers the almost unique possibility to print both the dielectric substrate and the metallic pattern. This possibility extends the possible future features implemented by the C-IRS maintaining the overall low cost of the final realization.

In summary, the current state-of-the-art in inkjet printing has demonstrated the feasibility of large scale fabrication and compatibility not only with metal nanoparticles but also with dielectrics and carbon materials (nanotubes or graphene) and, this represents a potentially disruptive technology as we move towards low-cost, large scale reconfigurable devices in the sub-THz range. 
These considerations indicate that C-IRS in the lower mmWave band is feasible, and the speed and low cost of the printing process make this technology suitable for massive production.

\section{Numerical V2V Results}\label{sect:numerical_results}

This section demonstrates the benefits of having C-RIS/C-IRS on the sides of vehicles in a highway V2V scenario. The results are twofold: first, we analyze the blockage probability of both the direct TxV-RxV link and the direct link assisted by one or more C-IRS/C-RIS relays; next, we assess the SNR distribution, again considering the direct link only and the combination of direct link and C-IRS/C-RIS as passive relays.

The considered multi-lane highway road segment is $500$ m long with $N_l=5$ lanes, each of $w_l = 5$ m width. The vehicles are randomly distributed on each lane according to a point Poisson process \cite{abul2007modeling} with a traffic density $\rho$ cars/km. All vehicles in the scenario have a rectangular occupation region of $(l_v \times w_v \times h_v) $ m and are equipped with two C-IRS/C-RIS on their left and right sides, as well as a transmitting/receiving ULA on the top side. Unless otherwise mentioned, the system and communication parameters used in the simulations are detailed in table \ref{tab:SimParam}.

\begin{table}[t!]
    \centering
    \caption{Simulation Parameters}
    \begin{tabular}{l|c|c}
    \toprule
        \textbf{Parameter} &  \textbf{Symbol} & \textbf{Value(s)}\\
        \hline
        Carrier frequency & $f$  & $28$ GHz \\
        Number of TxV/RxV antennas & $K$ & $8$\\
        Number of C-IRS/RIS elements & $M \times N$ & $400 \times 400$\\
        TxV/RxV element spacing & $d$ & $\lambda/2$\\
        C-IRS/C-RIS element spacing & $d_m, d_n$ & $\lambda/4$\\
        C-IRS/C-RIS curvature radius &$R$& $2,8$ m\\
        C-IRS configuration param. &$\overline{\theta}$ & $75$ deg\\
        Transmitted power & $\sigma^2_s$ & $10$ dBm\\
        Noise power & $\sigma^2_n$ & $-88$ dBm\\
        Vehicle shape & $l_v \times w_v \times h_v$ & $5 \times 1.8 \times 1.5$ m\\
        \bottomrule
    \end{tabular}
    \label{tab:SimParam}
\end{table}

\subsection{Relay Selection}
\begin{figure}[t!]
    \centering
    \includegraphics[width=0.5\columnwidth]{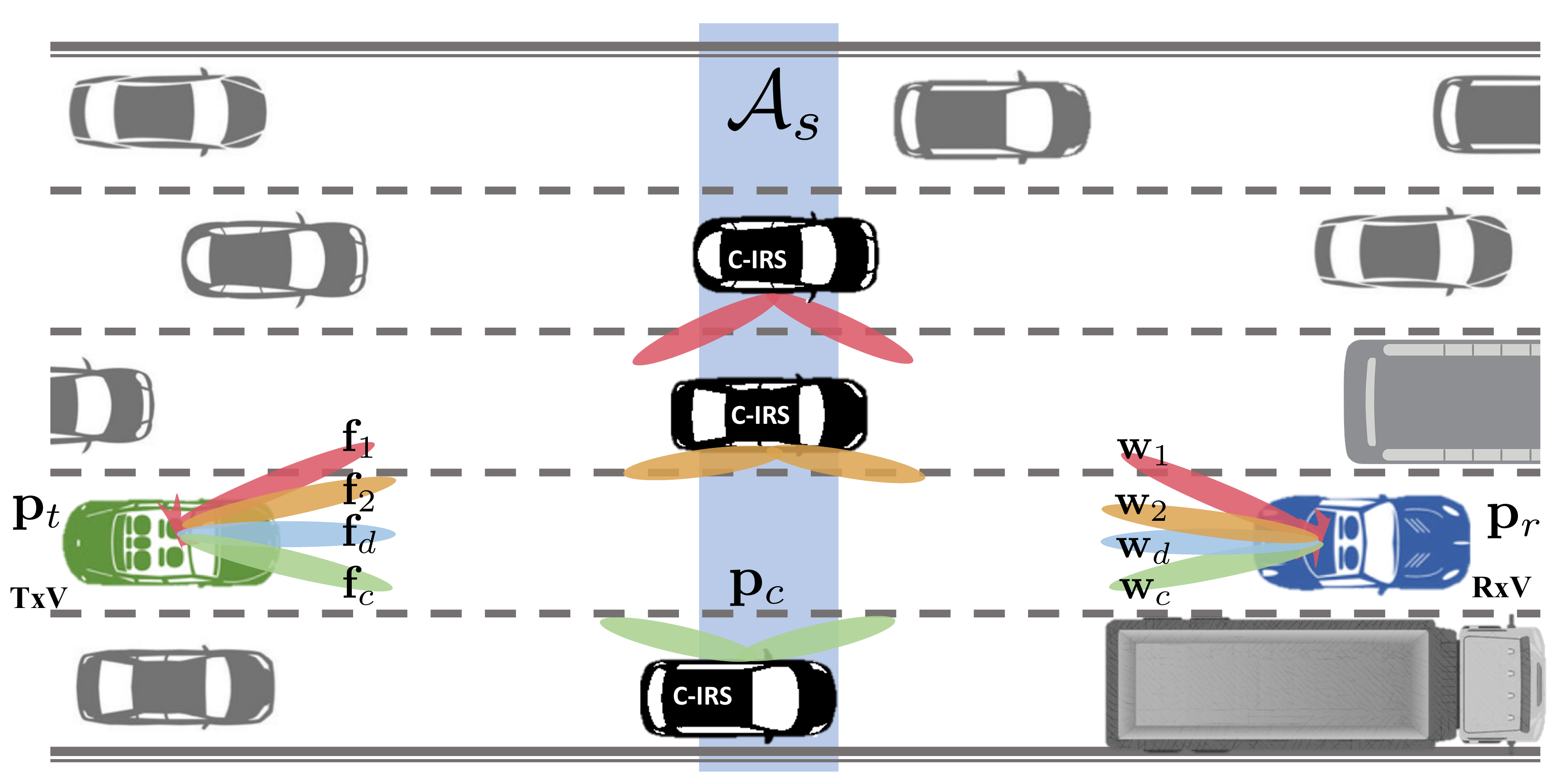}
    \caption{Example of C-IRS selection and specular area $\mathcal{A}_s$}
    \label{fig:As}
\end{figure}

Let us assume that the TxV and RxV know their positions, namely $\mathbf{p}_t$ and $\mathbf{p}_r$, as well as the position of neighboring CAVs, in which there is the $c$-th candidate passive relay, located in $\mathbf{p_c}$. For instance, position information can be obtained through signaling or sensing \cite{tagliaferri2021bbeampointing,Brambilla2020positioning}. When CAVs are equipped with C-RIS, finding the optimal passive relay in case of blockage of the direct link turns into a joint CAV selection \textit{and} C-RIS phase configuration that maximize the end-to-end SNR. In vehicular networks, this requires solving an optimization problem in real-time as the V2V scenario is rapidly time-varying. Therefore, selecting the optimal relaying C-RIS while configuring its phase (Section 	\ref{sect:reflection_conformal}) is computationally demanding as it requires high control signaling overhead. However, on the other hand, any CAV in range with TxV and RxV can serve as a passive relay. Differently, for passive and pre-configured C-IRS, the relay selection is considerably simpler, as the phases are fixed, yielding a lower number of potential relays. Since C-IRS are designed for specular reflections, potential relaying CAVs are possibly located halfway between TxV and RxV, within the \textit{specular area} $\mathcal{A}_s = W_s \times L_s$. Area $\mathcal{A}_s$ is rectangular and centered in $\mathbf{p}_s = (\mathbf{p}_r - \mathbf{p}_t)/2$, spanning all the highway lanes, as shown in Fig. \ref{fig:As}. While $W_s = N_l w_l$, $L_s$ is chosen to be two times the length of the C-IRS, i.e., $L_s = 2L$. If a CAV is inside the specular area, i.e., $\mathbf{p}_c \in \mathcal{A}_s$, then it can be selected as potential C-IRS relay for TxV-RxV V2V link.
The TxV and RxV beamformers $\mathbf{f}$ and $\mathbf{w}$ in \eqref{eq:receivedSignal} are chosen from beamforming codebooks $\mathcal{F}$ and $\mathcal{W}$ built upon the knowledge of the TxV, RxV and candidate relay positions within $\mathcal{A}_s$. Considering the position of the $c$-th relay, the beamforming vectors for the link reflected through the $c$-th C-IRS is computed from \eqref{eq:beamformer}: 
\begin{align}
    \mathbf{f}_c & = \left[1,...,e^{-j\pi(K-1)\cos\theta_t}\right]^\mathrm{T} \\
    \mathbf{w}_c & = \left[1,...,e^{-j\pi(K-1)\cos\theta_r}\right]^\mathrm{T}
\end{align}
for azimuth angles 
\begin{equation}
    \theta_t = \arctan\left(\frac{y_c - y_t}{x_c-x_t}\right),\qquad \theta_r = \arctan\left(\frac{y_r - y_c}{x_r-x_c}\right).
\end{equation}
Therefore, the beam codebooks are defined as 
\begin{equation}
    \mathcal{F} = \{\mathbf{f}_d,\{\mathbf{f}_c\}\}, \qquad  \mathcal{W} = \{\mathbf{w}_d,\{\mathbf{w}_c\}\}
\end{equation}
where $\mathbf{f}_d$ and $\mathbf{w}_d$ denote the beamforming vectors for the direct TxV-RxV link.
Beamforming vectors $\mathbf{f}$ and $\mathbf{w}$ are then obtained by selecting among the set of available beamformers $\mathcal{F}$ and $\mathcal{W}$, the tuple that maximizes the received power, i.e.,
\begin{equation}\label{eq:optimalBeam}
    \mathbf{f}_{opt}, \mathbf{w}_{opt} = \underset{\substack{\mathbf{f}\in\mathcal{F}\\\mathbf{w}\in\mathcal{W}}}{\mathrm{argmax}}\,  \big\lvert\mathbf{w}^\mathrm{H}\mathbf{H}\mathbf{f} \,s + \mathbf{w}^\mathrm{H} \mathbf{n}\big\rvert^2.
\end{equation}
\begin{figure}[tb!]
    \centering
    \includegraphics[width=0.5\columnwidth]{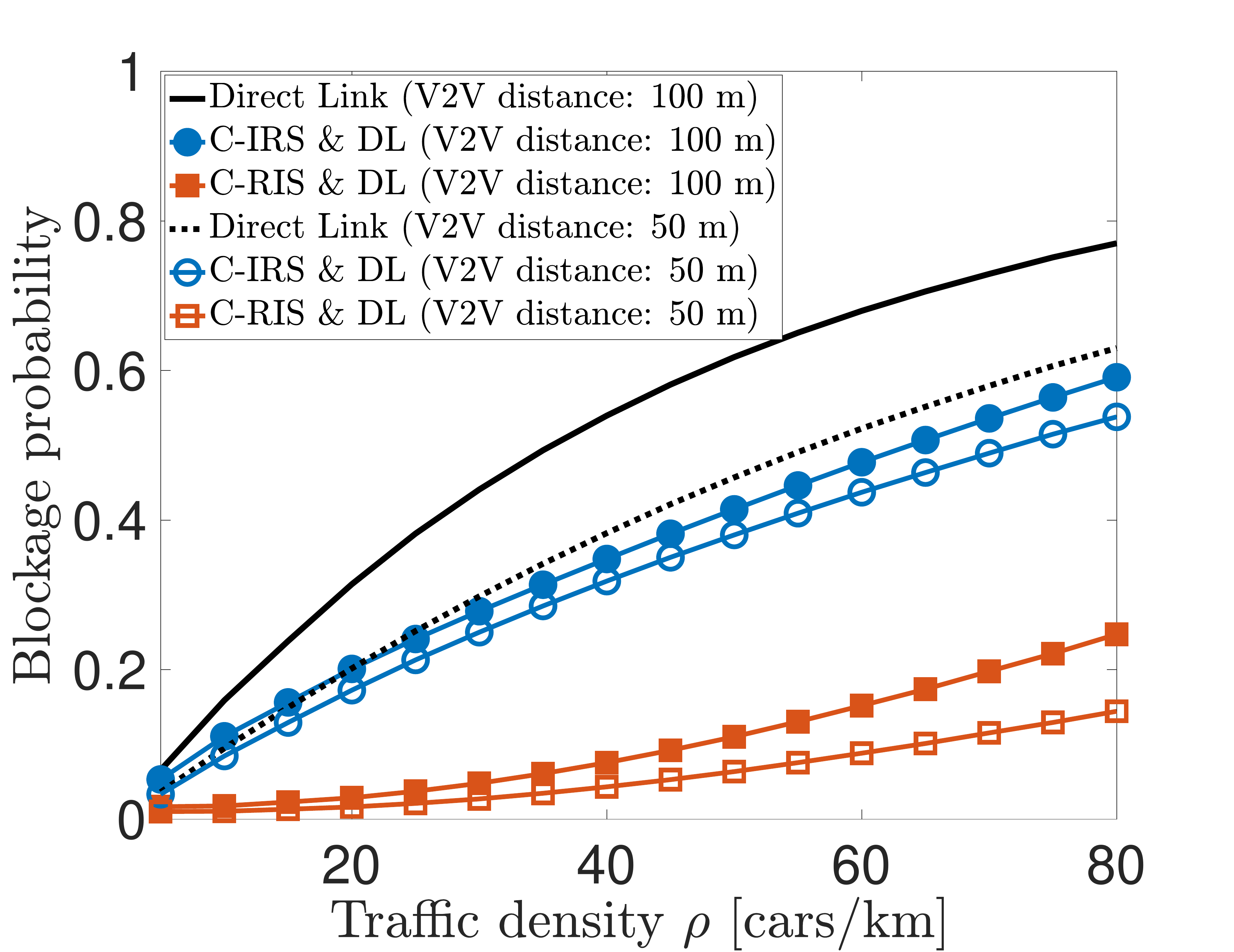}
    \caption{Blockage probability varying the traffic density}
    \label{fig:BP}
\end{figure}

\subsection{Results}

Figure \ref{fig:BP} shows the remarkable benefits provided by employing conformal metasurfaces on car doors. The blockage probability of the V2V link is reported in Fig. \ref{fig:BP}, varying the traffic density $\rho$, for a TxV-RxV distance of $r_d =50$ m and $r_d =100$ m, when using \textit{(i)} the direct link only (black lines); \textit{(ii)} the best link between the direct one and the pre-configured C-IRS-enabled one, according to \eqref{eq:optimalBeam} (blue marked lines); \textit{(iii)} the best link between direct and C-RIS-enabled one (red marked lines). The latter assumes an instantaneous real-time phase reconfiguration of the C-RIS according to a perfect CSI knowledge. For C-IRS, the configuration parameter $\overline{\theta}$ controlling the angular bandwidth of specular reflection is set to $75$ deg, according to the distribution in Fig. \ref{subfig:AZpdf}. As depicted in Fig. \ref{fig:BP}, the blockage probability of the direct links (black lines) increase with traffic density, while we observe a substantial reduction in blockage probability of $\approx 20\%$ when using C-IRS combined with the direct link and of $\approx 70\%$ when using C-RIS and the direct link. In the second case, however, C-RIS-equipped CAVs do not need to be in $\mathcal{A}_s$ to serve as relays. Therefore, the average number of relaying CAVs equipped by C-RIS is higher than in case of C-IRS, justifying the improved performance, at the price of an intensive control signaling for the relay link setup. 

\begin{figure*}[!t]
    \centering
    \subfloat[][$R=2$ m, $r_d=50$ m]{\includegraphics[width=0.45\columnwidth]{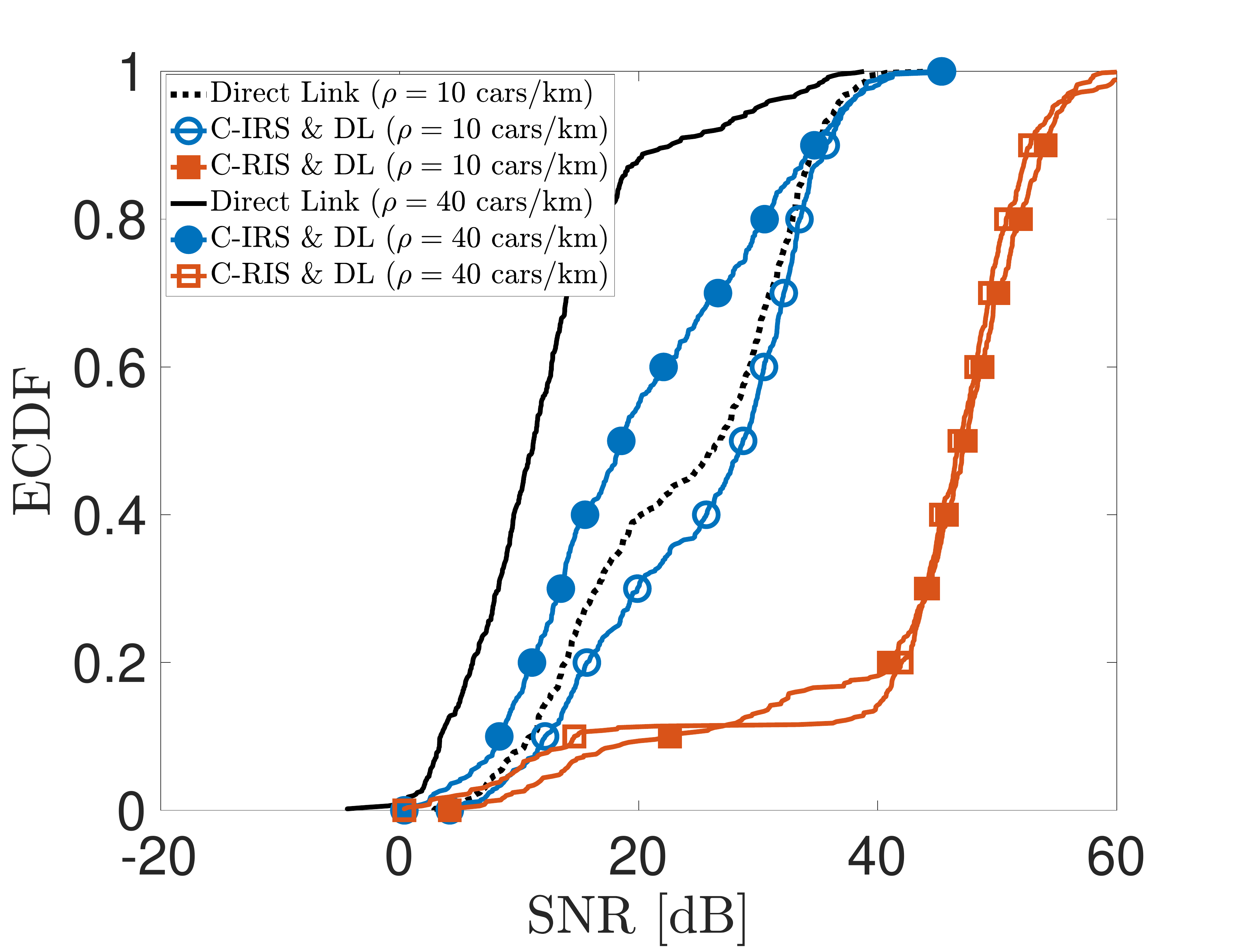}\label{subfig:ECDF_d50_R2_SNR_TD10}}
    \subfloat[][$R=2$ m, $r_d=100$ m]{\includegraphics[width=0.45\columnwidth]{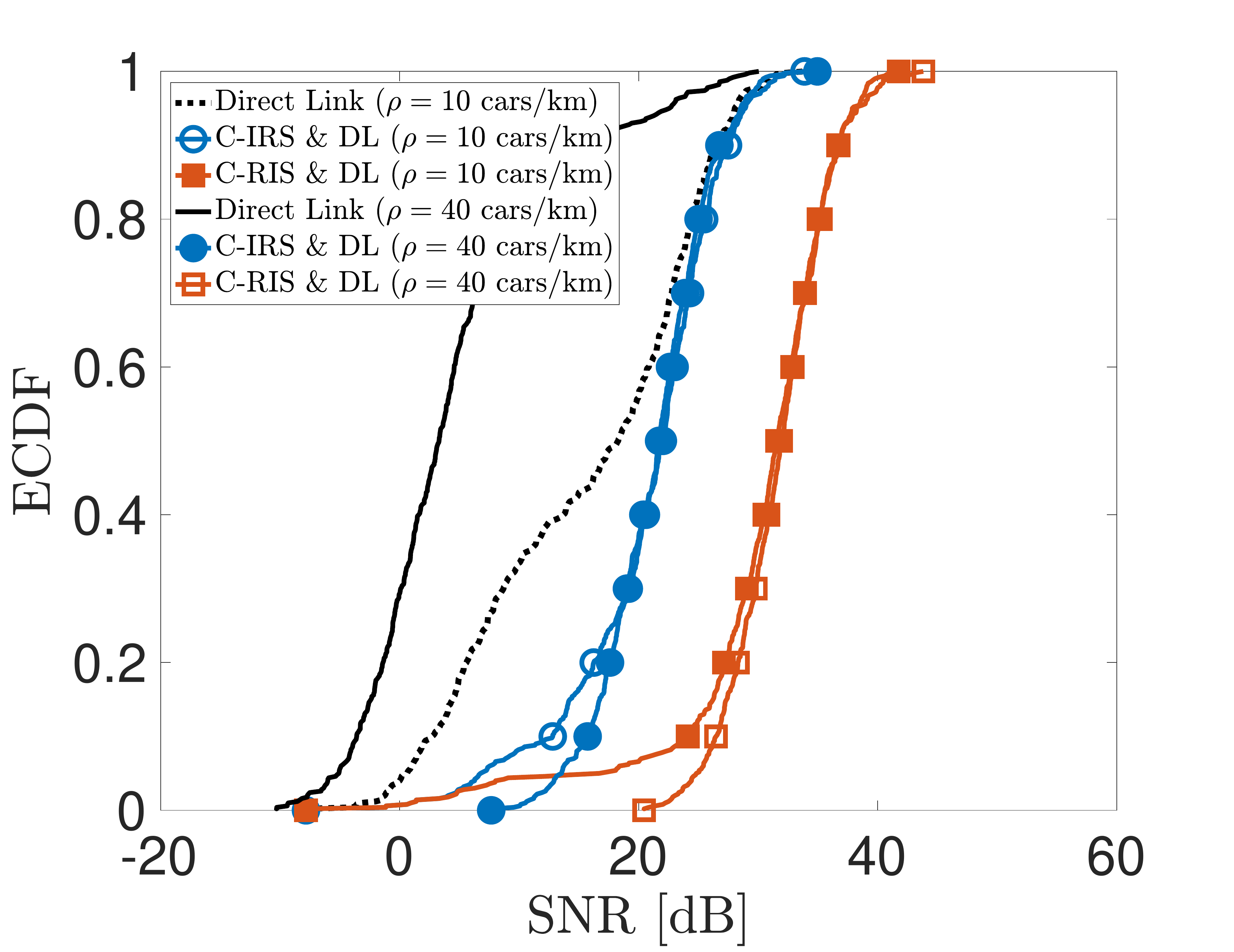}\label{subfig:ECDF_d100_R2_SNR_TD10}}\\
    \subfloat[][$R=8$ m, $r_d=50$ m]{\includegraphics[width=0.45\columnwidth]{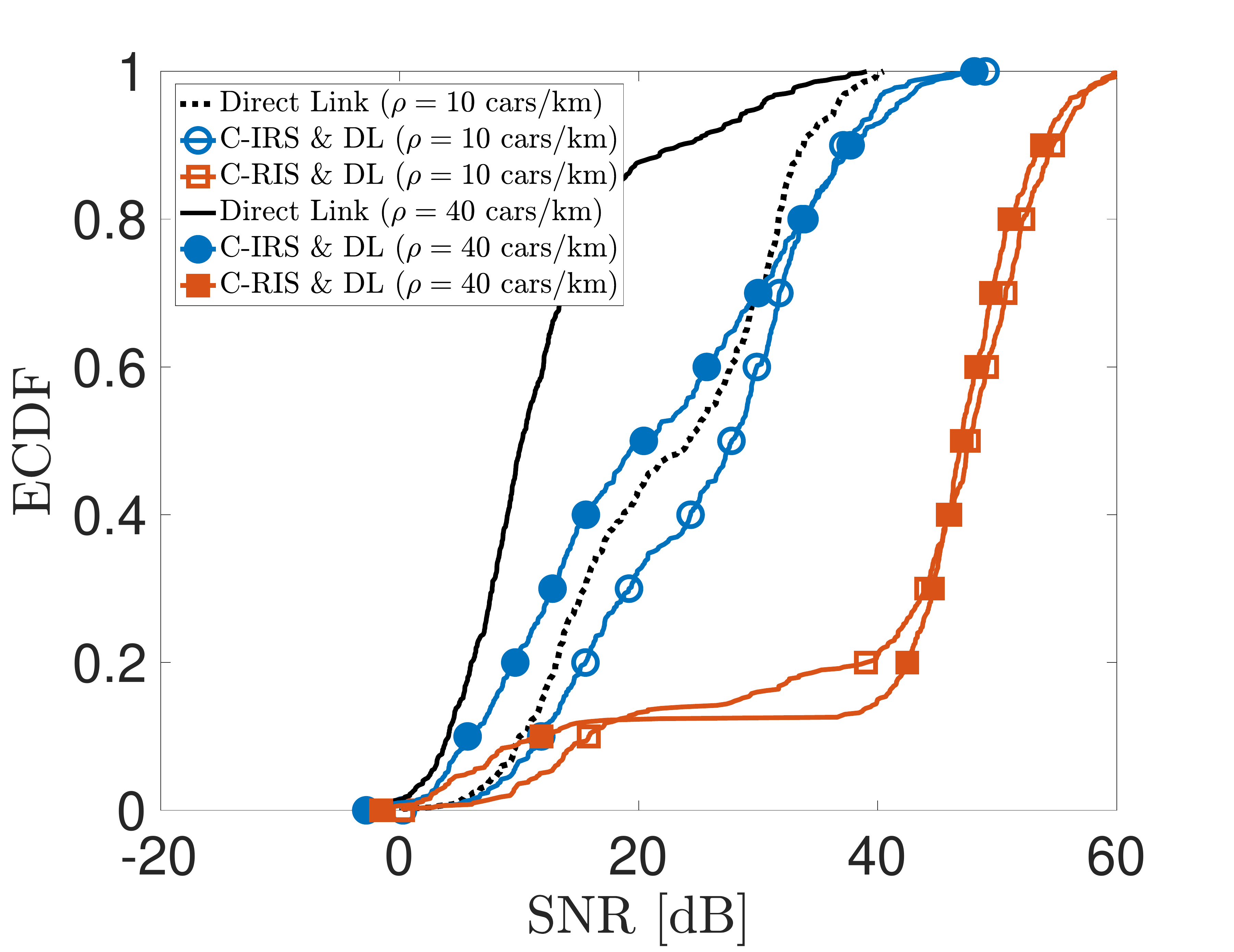}\label{subfig:ECDF_d50_R8_SNR_TD10}}
    \subfloat[][$R=8$ m, $r_d=100$ m]{\includegraphics[width=0.45\columnwidth]{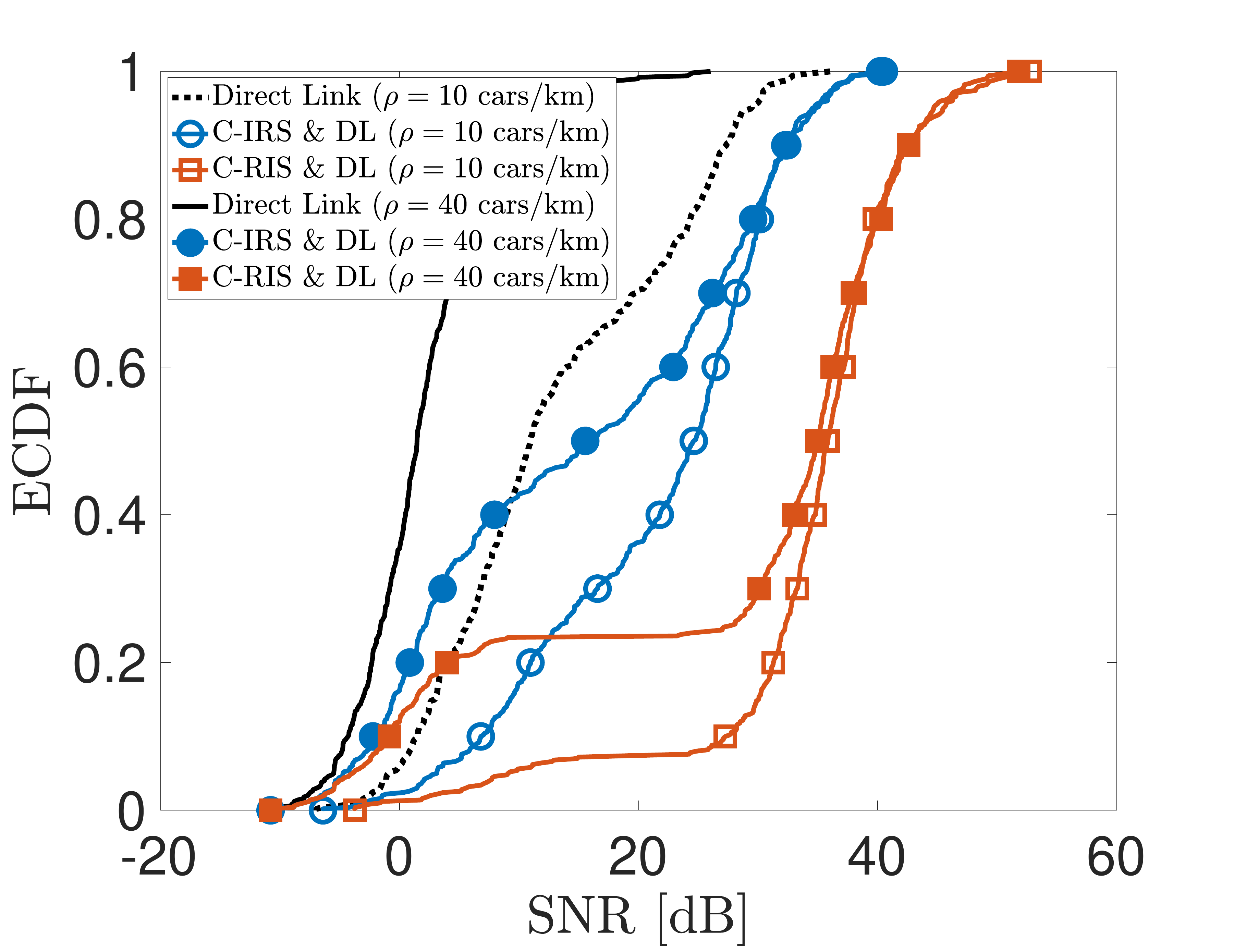}\label{subfig:ECDF_d100_R8_SNR_TD10}}\\
    \caption{ECDF of the SNR for (\ref{subfig:ECDF_d50_R2_SNR_TD10},\ref{subfig:ECDF_d50_R8_SNR_TD10}) V2V distance $r_d=50$ m and (\ref{subfig:ECDF_d100_R2_SNR_TD10},\ref{subfig:ECDF_d100_R8_SNR_TD10}) $40$ cars/km (on each lane). }
    \label{fig:ECDF_SNR}
\end{figure*}
The second set of results is focused on the empirical cumulative distribution function (ECDF) of the SNR at the RxV, defined as
\begin{equation}\label{eq:SNR}
    \mathrm{SNR} = \frac{\sigma^2_s\lvert\mathbf{w}^\mathrm{H}\mathbf{H}\mathbf{f}\rvert^2}{K\sigma^2_n}.
\end{equation}
The CDFs are in Figs. \ref{fig:ECDF_SNR}, varying the radius of curvature $R$ of the C-RIS/C-IRS and two values of traffic density: medium-to-low traffic ($\rho=10$ cars/km) and high traffic ($\rho=40$ cars/km). The general trend of the curves highlights the remarkable performance improvement of using either C-RIS or C-IRS, compared to a system relying only of the direct V2V link. On average, we observe that when the TxV-RxV distance is $r_d=50$ m, the usage of pre-configured fully passive C-IRS provides an SNR improvement of $\approx 3$ dB in medium-to-low traffic density and $\approx 10$ dB for severe traffic. C-RIS SNR gain is instead $\approx 20$ dB for $\rho=10$ cars/km and in excess of $35-40$ dB for $\rho=40$ cars/km. When the TxV-RxV distance $r_d$ increases to $100$ m, the blockage affection becomes more severe and the impact of relay links dominate the performance of the overall V2V SNR, especially in high traffic conditions. The SNR gain provided by C-IRS compared to the direct link increases by $15-20$ dB, while no difference is observed for C-RIS. Thus, interestingly, the relative C-RIS-to-C-IRS performance gap diminishes. According to these results, two observations can be made. First, we can notice that C-RIS are not substantially affected by traffic density. Regardless of the TxV-RxV, the probability of finding at least one relaying vehicle does not appreciably change with $\rho$, thanks to the possibility of using any CAV as a relay with a proper phase configuration (in principle, for any incidence/reflection angles).
A second noticeable observation is that, in practice, the effect of the curvature $R$ is negligible, as the V2V link does not experience any improvement passing from $R=2$ m to $R=8$ m. This shows that a proper pre-configuration of the C-IRS, by means of parameter $\overline{\theta}$, allows to match the experienced azimuth angles in the vehicular network without appreciable performance reduction.

\section{Conclusions and Open Challenges} \label{sect:conclusion}

With the advent of self-driving cars, vehicles are expected to share a massive amount of data with neighbors to augment the environment perception for safety-critical applications or onboard infotainment. In this setting, mmWave/sub-THz bands arose as the most promising solution to guarantee the required data rates through directive V2V links. As spectrum increases the blockage from random vehicles drastically reduces the reliability of V2V links, calling for advanced blockage countermeasures. This paper proposes the usage of conformal-designed metasurfaces deployed on CAVs' doors as passive relay in mmWave/sub-THz vehicular networks. Since vehicles' doors are, in general, complex conformal surfaces that cannot be considered EM flat, we first analytically derived the optimal phase pattern to be applied to arbitrarily shaped (conformal) RIS (C-RIS) given the exact knowledge of incident and reflection angles. Then, we remove the latter requirement by proposing a novel phase design for non-reconfigurable fully passive C-IRS, where metasurfaces act as mirrors compensating for the non-flat door's shape. Numerical simulations illustrate the advantages of using the combination of a direct link and C-RIS/C-IRS-enabled relays in a multi-lane highway vehicular scenario, compared to direct V2V links only. C-RIS provides the best overall performance improvement in terms of blockage probability reduction ($-70\%$) and SNR ($+30$ dB), at the price of an intensive control signaling for real-time phase reconfiguration, as well as the perfect CSI knowledge at the C-RIS. By contrast, C-IRS performance is slightly lower ($-20\%$ of blockage probability and $+15$ dB of average SNR); however, the relay-assisted V2V link can operate without control signaling except for the relay selection. These results justify a possible adoption of C-RIS and C-IRS to reduce the blockage affection in future 6G vehicular networks, provided that their cost is negligible for car parts manufacturing. Among the open challenges, the most relevant is the characterization of the augmentation of mutual vehicle interference generated by a certain high spatial density of C-RIS/C-IRS relay links, with design of proper countermeasures. Moreover, the SNR gain gap between the C-RIS and C-IRS solutions can reach 20dB. This suggests that practical solutions can be investigated for C-IRS to narrow this gap with little or no reconfigurability.

\section*{Acknowledgment}
The work has been partially supported by the Huawei-Politecnico di Milano Joint Research Lab. 

\bibliographystyle{IEEEtran}
\bibliography{biblio}

\end{document}